\documentclass[sigconf]{acmart}

\usepackage{microtype}
\usepackage{multirow}
\usepackage[normalem]{ulem}
\usepackage{graphicx}
\usepackage{blindtext}
\usepackage{subfiles}
\usepackage{subcaption}
\usepackage{float}
\usepackage{afterpage}
\usepackage{color}
\usepackage{enumitem}
\usepackage{placeins}
\usepackage{float}
\usepackage{colortbl}
\usepackage[table]{xcolor} 
\definecolor{lightgray}{gray}{0.9} 
\usepackage{amsmath}
\usepackage{amsthm}
\usepackage{mathtools}
\hypersetup{colorlinks=true, linkcolor=blue, urlcolor=blue, citecolor=blue}

\makeatletter
\@ifundefined{Bbbk}{}{}
\makeatother
\usepackage{amssymb}

\usepackage{booktabs} 

\newcommand{\E}{\mathbb{E}}
\newcommand{\R}{\mathbb{R}}
\newcommand{\setU}{\mathcal{U}}
\newcommand{\setI}{\mathcal{I}}
\newcommand{\setS}{\mathcal{S}}

\newcommand{\vu}{\mathbf{e}_u}
\newcommand{\vi}{\mathbf{e}_i}
\newcommand{\vj}{\mathbf{e}_j}

\newcommand{\vuorig}{\mathbf{e}_{u,orig}}

\newcommand{\vecv}{\mathbf{v}}

\newcommand{\popdir}{\mathbf{e}_{pop}}
\newcommand{\predir}{\mathbf{e}_{pref}}

\newcommand{\dopt}{\mathbf{d}_{u}^{*}}
\newcommand{\gpos}{\mathbf{g}_{u}^{pos}}
\newcommand{\gneg}{\mathbf{g}_{u}^{neg}}

\newcommand{\defeq}{\vcentcolon=}
\newcommand{\norm}[1]{\left\lVert#1\right\rVert}

\newtheorem{theorem}{Theorem}[section]
\newtheorem{lemma}[theorem]{Lemma}
\newtheorem{proposition}[theorem]{Proposition}

\newtheorem{definition}{Definition}[section]
\newtheorem{assumption}{Assumption}[section]

\useunder{\uline}{\ul}{}
\graphicspath{images/}
\AtBeginDocument{%
  }

\copyrightyear{2026}
\acmYear{2026}
\setcopyright{cc}
\setcctype{by}
\acmConference[KDD '26]{Proceedings of the 32nd ACM SIGKDD Conference on Knowledge Discovery and Data Mining V.1}{August 09--13, 2026}{Jeju Island, Republic of Korea}
\acmBooktitle{Proceedings of the 32nd ACM SIGKDD Conference on Knowledge Discovery and Data Mining V.1 (KDD '26), August 09--13, 2026, Jeju Island, Republic of Korea}
\acmPrice{}
\acmDOI{10.1145/3770854.3780295}
\acmISBN{979-8-4007-2258-5/2026/08}

\title{Rethinking Popularity Bias in Collaborative Filtering via Analytical Vector Decomposition}

\author{Lingfeng Liu}
\affiliation{
\institution{School of Artificial Intelligence and Data Science, University of Science and Technology of China}
\city{Hefei}
\country{China}
}\email{lfliu@mail.ustc.edu.cn}

\author{Yixin Song}
\affiliation{
\institution{School of Artificial Intelligence and Data Science, University of Science and Technology of China}
\city{Hefei}
\country{China}
}\email{yixinsong@mail.ustc.edu.cn}

\author{Dazhong Shen}
\affiliation{
\institution{College of Computer Science and Technology, Nanjing University of Aeronautics and Astronautics}
\city{Nanjing}
\country{China}
}\email{shendazhong@nuaa.edu.cn}

\author{Bing Yin}
\affiliation{
  \institution{State Key Laboratory of Cognitive Intelligence \\ iFLYTEK Research, iFLYTEK}
  \city{Hefei}
  \country{China}
}\email{bingyin@iflytek.com}

\author{Hao Li}
\affiliation{
\institution{iFLYTEK Research, iFLYTEK}
\city{Hefei}
\country{China}
}\email{haoli5@iflytek.com}

\author{Yanyong Zhang}
\affiliation{
\institution{School of Artificial Intelligence and Data Science, University of Science and Technology of China}
\city{Hefei}
\country{China}
}\email{yanyongz@ustc.edu.cn}

\author{Chao Wang}
\authornote{Chao Wang is the corresponding author.}
\affiliation{
\institution{School of Artificial Intelligence and Data Science, University of Science and Technology of China \\ State Key Laboratory of Cognitive Intelligence}
\city{Hefei}
\country{China}
}
\email{wangchaoai@ustc.edu.cn
}
                             
\renewcommand{\shortauthors}{Lingfeng Liu et al.}

\usepackage{tcolorbox}

\begin{document}

\begin{abstract}
Popularity bias fundamentally undermines the personalization capabilities of collaborative filtering (CF) models, causing them to disproportionately recommend popular items while neglecting users' genuine preferences for niche content. While existing approaches treat this as an external confounding factor, we reveal that popularity bias is an intrinsic geometric artifact of Bayesian Pairwise Ranking (BPR) optimization in CF models. Through rigorous mathematical analysis, we prove that BPR systematically organizes item embeddings along a dominant ``popularity direction'' where embedding magnitudes directly correlate with interaction frequency. This geometric distortion forces user embeddings to simultaneously handle two conflicting tasks—expressing genuine preference and calibrating against global popularity—trapping them in suboptimal configurations that favor popular items regardless of individual tastes. We propose \textbf{D}irectional \textbf{D}ecomposition and \textbf{C}orrection (\textbf{DDC}), a universally applicable framework that surgically corrects this embedding geometry through asymmetric directional updates. DDC guides positive interactions along personalized preference directions while steering negative interactions away from the global popularity direction, disentangling preference from popularity at the geometric source. Extensive experiments across multiple BPR-based architectures demonstrate that DDC significantly outperforms state-of-the-art debiasing methods, reducing training loss to less than 5\% of heavily-tuned baselines while achieving superior recommendation quality and fairness. Code is available in \url{https://github.com/LingFeng-Liu-AI/DDC}.
\end{abstract}



\maketitle

\newcommand\kddavailabilityurl{https://doi.org/10.5281/zenodo.18048217} 
\ifdefempty{\kddavailabilityurl}{}{
\begingroup\small\noindent\raggedright\textbf{Resource Availability:}\\
The source code of this paper has been made publicly available at \url{\kddavailabilityurl}.
\endgroup
}

\section{Introduction}
\label{sec:intro}
\begin{figure}[t]
    \centering
    \includegraphics[width=1.0\linewidth]{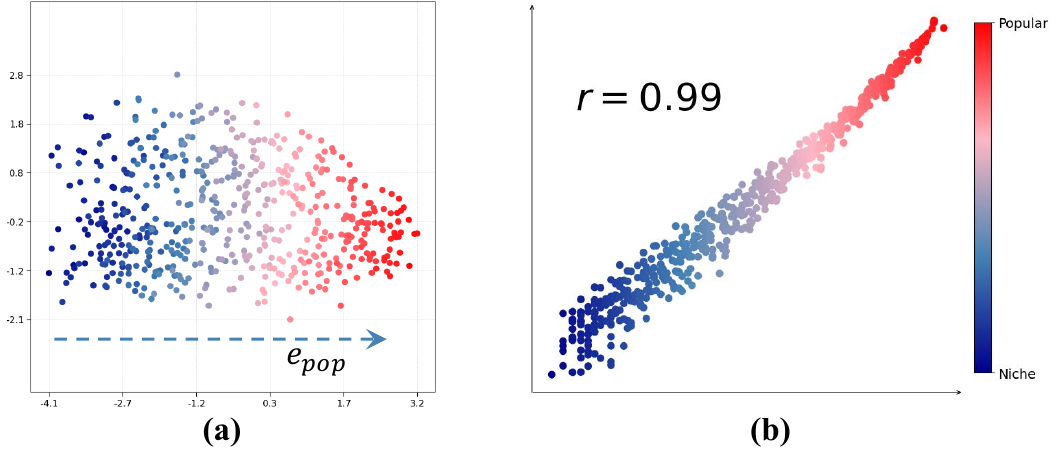} 
    \caption{
        \textbf{Geometric visualization and quantification of popularity bias in BPR-based CF models.}
        \textbf{(a)} The 2D projection of item embeddings from a BPR-based CF model. Embeddings are structurally organized along a dominant popularity direction ($\popdir$).
        \textbf{(b)} The projection magnitude of item embeddings onto $\popdir$ exhibits a near-perfect linear correlation (Pearson's $r=0.99$) with their actual popularity, quantitatively confirming the geometric bias.
    }
    \label{fig:intropic}
\end{figure}

Collaborative filtering (CF), which models user preferences based on historical interactions, serves as the cornerstone of modern recommender systems~\cite{DisentanglingUserZheng2021,ImprovingRecommendationChen2023,ASurveyonAccuracyWu2023,wang2018confidence,wang2020setrank}. Embedding-based models, particularly those leveraging graph neural networks like LightGCN~\cite{LightGCNHe2020}, have achieved great success by learning user and item representations in a latent space~\cite{GraphNeuralWu2023,wang2021variable,wang2021personalized}.
These models are almost universally trained with the Bayesian Pairwise Ranking (BPR) loss, a de facto standard for implicit feedback datasets. 
However, BPR optimization exhibits a fundamental geometric flaw: it systematically distorts the embedding space to favor popular items, creating a popularity bias that undermines personalization~\cite{BiasandDebiasChen2023,PopularityBiasZhu2021}.
This bias not only reduces recommendation personalization and diversity~\cite{AutoDebiasChen2021,ItemSideJiang2024,FairCFShao2022}, but also perpetuates a Matthew effect where popular items dominate while niche content is systematically overlooked~\cite{ASurveyontheFairnesWang2023,zhu2024graph,wang2025face}.

Traditional approaches in recommender systems typically view popularity bias as an external confounding factor that interferes with recommendation quality, rather than recognizing it as an intrinsic consequence of the optimization process itself. These methods employ causal inference or additional bias modeling techniques to separately address bias effects without integrating deeply with the embedding learning mechanism. For example, re-weighting methods, such as inverse propensity scoring (IPS), attempt to balance training data by down-weighting popular item interactions~\cite{CounterfactualReasoningBottou2013, AutoDebiasChen2021}. Regularization approaches penalize correlations between recommendation scores and item popularity~\cite{IncorporatingDiversityJacek2016, ControllingPopularityAbdollahpouri2017}. Recently, causal methods model and remove confounding effects~\cite{DeconfoundedRecommendationWang2021, CausalInterventionZhang2021} through causal analyses.
However, these macroscopic approaches address symptoms rather than root causes~\cite{ModelAgnosticWei2021}. They fail to answer a fundamental question: how does BPR optimization geometrically distort the latent representation space to systematically favor popular items? Understanding this geometric mechanism is crucial for developing principled solutions that address bias at its source.

Differently from these surface-level interventions, this paper directly confronts the foundational gap of understanding how BPR optimization geometrically creates popularity bias in the embedding space itself. We ask two key research questions: First, can we identify the precise geometric mechanism through which BPR systematically favors popular items? Second, can this understanding enable a principled, model-agnostic correction method that universally applies to existing BPR-based recommenders?
Our core finding is the identification of a dominant ``popularity direction'', $\popdir$, in the embedding space of BPR-based CF models. Taking LightGCN trained on MovieLens-1M as a representative example, as visualized and quantified in Figure~\ref{fig:intropic}, we provide empirical evidence that an item's popularity is directly and strongly correlated with the magnitude of its embedding's projection onto this single direction. This geometric alignment reveals that BPR systematically conflates an item's intrinsic qualities with its global popularity, trapping user representations in suboptimal configurations that favor popular items regardless of individual preferences.

Based on this discovery, we propose \textbf{D}irectional \textbf{D}ecomposition and \textbf{C}orrection (\textbf{DDC}), a novel and lightweight framework designed to rectify the distorted embedding geometry in BPR-based CF models. The key advantage of DDC is its broad applicability—it can transform existing BPR-trained recommenders through a simple fine-tuning stage, without requiring architectural modifications or model-specific adaptations.
Through rigorous geometric analysis, we mathematically derive how BPR's pairwise ranking objective inevitably creates a dominant popularity direction in the embedding space, where item embeddings align along this direction with magnitudes proportional to their popularity. We further prove that this geometric distortion causes BPR gradients to become misaligned with the ideal update direction, forcing user embeddings to simultaneously handle two conflicting tasks: expressing genuine preference and calibrating against global popularity. Based on this theoretical foundation, we design DDC with an asymmetric two-directional correction strategy: (1) a personalized preference direction that guides positive updates toward authentic user interests, and (2) a global popularity direction that steers negative updates away from popularity-driven patterns. This directional decomposition surgically disentangles preference from popularity at the geometric source of the bias.
Extensive experiments on three benchmark datasets demonstrate that DDC significantly outperforms state-of-the-art debiasing methods while achieving dramatically accelerated convergence across different BPR-based architectures. Notably, our method reduces the final BPR loss on training data to less than 5\% of heavily-tuned baselines (see Section~\ref{sec:conv}), providing strong evidence that DDC guides user embeddings toward more accurate representations of true preference.
Our contributions are threefold:
\begin{itemize}[leftmargin=*]
\item We identify and theoretically characterize the dominant popularity direction in BPR embedding spaces, revealing the geometric root of popularity bias through rigorous mathematical analysis.
\item We propose DDC, a universally applicable framework that transforms existing BPR-based CF models through asymmetric two-directional corrections without architectural changes.
\item We provide comprehensive experimental validation across multiple BPR architectures, showing that DDC achieves superior debiasing performance through fundamental geometric correction, not by increasing model complexity.
\end{itemize}

\section{Preliminary}
\label{sec:preliminary}
In this section, we introduce the problem setting, foundational CF models and the Bayesian Pairwise Ranking loss.

\subsection{Problem Formulation}
Let $\mathcal{U}$ and $\mathcal{I}$ denote the sets of users and items, respectively. We are given a set of historical user-item interactions $\mathcal{R} \subseteq \mathcal{U} \times \mathcal{I}$, which represent implicit feedback. If a pair $(u, i) \in \mathcal{R}$, it signifies that user $u$ has interacted with item $i$. The vast majority of user-item pairs are unobserved. The core task of collaborative filtering is to predict a personalized ranked list of items from the unobserved set for each user $u \in \mathcal{U}$.
Embedding-based models tackle this by learning low-dimensional representations for users and items. Specifically, each user $u \in \mathcal{U}$ and item $i \in \mathcal{I}$ are mapped to embedding vectors $\mathbf{e}_u, \mathbf{e}_i \in \mathbb{R}^d$, where $d$ is the latent dimension. The preference score, reflecting the similarity between a user and an item, is then computed from their embeddings.

\subsection{CF Models}
Our investigation focuses on the geometric properties of CF embeddings. Here, we introduce two representative models: 

\subsubsection{Matrix Factorization (MF)}
Matrix Factorization~\cite{BPRRendle2009} is a foundational collaborative filtering model. It directly learns the user and item embeddings, $\mathbf{e}_u$ and $\mathbf{e}_i$, from the interaction data. The preference score $\hat{y}_{ui}$ that user $u$ is predicted to have for item $i$ is calculated as the inner product of their respective embeddings:
\begin{equation}
\hat{y}_{ui} = \mathbf{e}_u^T \mathbf{e}_i.
\end{equation}
\subsubsection{LightGCN}
LightGCN~\cite{LightGCNHe2020} is a state-of-the-art GNN-based model that simplifies the design of traditional Graph Convolutional Networks (GCNs) for recommendation. 
LightGCN begins with initial, learnable embeddings $\mathbf{E}^{(0)} = [\mathbf{e}_{u_1}^{(0)}, ..., \mathbf{e}_{i_1}^{(0)}, ...]^T$ and refines them through a linear propagation mechanism on the user-item interaction graph. At each layer $k$, the embeddings are aggregated from their neighbors as follows:
\begin{equation}
\mathbf{E}^{(k)} = (\mathbf{D}^{-1/2}\mathbf{A}\mathbf{D}^{-1/2}) \mathbf{E}^{(k-1)},
\end{equation}
\noindent where $\mathbf{A}$ is the adjacency matrix of the user-item graph and $\mathbf{D}$ is the diagonal degree matrix. The final representation for each user and item is typically a weighted sum of the embeddings from all layers:
$\mathbf{e}_u = \sum_{k=0}^{K} \alpha_k \mathbf{e}_u^{(k)}  \text{ and } \mathbf{e}_i = \sum_{k=0}^{K} \alpha_k \mathbf{e}_i^{(k)},$
where $K$ is the number of layers and $\alpha_k$ are layer-wise combination weights. Critically, despite its more sophisticated representation learning process, LightGCN ultimately computes the final preference score using the same inner product as MF: $\hat{y}_{ui} = \mathbf{e}_u^T \mathbf{e}_i$. This shared final step is key to understanding why the bias we investigate is prevalent across different model architectures.

\subsection{Bayesian Pairwise Ranking (BPR) Loss}
The Bayesian Pairwise Ranking (BPR) loss is the standard optimization objective for models trained on implicit feedback datasets~\cite{BPRRendle2009}. It is built on a pairwise learning assumption: for a given user, an observed item should be ranked higher than all unobserved items. To operationalize this, BPR constructs training instances as triplets $(u, i, j)$, where $u$ is a user, $i$ is an item they have interacted with (a positive item, $(u, i) \in \mathcal{R}$), and $j$ is an item they have not interacted with (a negative item, $(u, j) \notin \mathcal{R}$).
The objective is to maximize the score difference between the positive and negative items, $\hat{y}_{ui} - \hat{y}_{uj}$. The BPR loss function is formulated as:

\begin{equation}
\mathcal{L}_{BPR} = \sum_{(u,i,j) \in \mathcal{D}} -\ln \sigma(\mathbf{e}_u^T \mathbf{e}_i - \mathbf{e}_u^T \mathbf{e}_j) + \lambda ||\Theta||_2^2.
\end{equation}
where $\mathcal{D}$ is the set of all training triplets, $\sigma(\cdot)$ is the sigmoid function, $\Theta$ represents all learnable model parameters (the embeddings $\mathbf{E}^{(0)}$ in LightGCN), and $\lambda$ is the L2 regularization hyperparameter. 

This mechanism of pushing user embeddings $\mathbf{e}_u$ closer to positive items $\mathbf{e}_i$ than negative items $\mathbf{e}_j$ systematically introduces the geometric distortion we investigate.

\section{Analysis of Popularity Bias}
\label{sec:analysis}
Our investigation reveals that the geometry of embeddings learned via BPR is not arbitrary but is systematically distorted by popularity. This section deconstructs this phenomenon mathematically, identifying it as the root cause of suboptimal model performance.

Due to page limits, we present the key insights here. Detailed theoretical analysis and proofs are in the Appendix~\ref{sec:appendix_theory}.

\subsection{The Geometric Imprint of Popularity}
\label{sec:imprint_of_popularity}
We begin by the definition of popularity in recommendations.
\begin{definition}[Item Popularity]
The popularity of an item $i$, denoted $Pop(i)$, is its interaction frequency in the training data, i.e., the size of the set of users who have interacted with it: $Pop(i) \defeq |\setU_i^+|$.
\end{definition}

Empirically, we find item embeddings are not isotropically distributed but are organized along a dominant axis correlated with item popularity. We identify this axis by computing the difference vector between the centroid of high-popularity items and that of low-popularity items.
Formally, let $\setI_{head}$ and $\setI_{tail}$ denote the sets of items with the highest and lowest interaction frequencies according to a predefined ratio $\rho$ (e.g., $\rho=0.05$). The popularity direction $\popdir$ is defined as the normalized difference between their centroids:
\begin{equation}
\label{eq:pop_dir_def}
\mbox{\footnotesize $\displaystyle
\popdir \defeq \frac{\vecv_{diff}}{\norm{\vecv_{diff}}}, \quad \text{where } \vecv_{diff} = \frac{1}{|\setI_{head}|}\sum_{i \in \setI_{head}}\vi - \frac{1}{|\setI_{tail}|}\sum_{i \in \setI_{tail}}\vi.
$}
\end{equation}
An item's popularity is strongly and positively correlated with the projection of its embedding onto this direction: $Pop(i) \propto \vi^T \popdir$.

This geometric structure is an inherent artifact of BPR optimization. To understand why, consider the gradient update for an item embedding $\vi$. For each positive interaction $(u,i)$, the BPR loss gradient pulls $\vi$ towards the user embedding $\vu$. The total expected update for item $i$ is driven by the sum of embeddings of all users who interacted with it:
\begin{equation}
\label{eq:item_grad_intuition}
\E[\Delta \vi] \propto \sum_{u \in \setU_i^+} \vu = Pop(i) \cdot \E_{u \in \setU_i^+}[\vu].
\end{equation}
The key insight is how the term $\E_{u \in \setU_i^+}[\vu]$ behaves for items of different popularity levels:
\begin{itemize}[leftmargin=*]
    \item \textbf{Popular Items}: A popular item $i_{pop}$ is liked by a large, diverse set of users. By the Law of Large Numbers, the average embedding of these users, $\E_{u \in \setU_i^+}[\vu]$, closely approximates the global average user embedding, $\E_{u \in \setU}[\vu]$. Hence, popular items are consistently pulled in this same, stable, mean-user direction.
     \item \textbf{Niche Items}: A niche item $i_{niche}$ is liked by a small, specific group of users. Their average embedding is idiosyncratic and points in a unique direction, not aligned with the global average.
\end{itemize}

Consequently, popular items receive powerful and directionally consistent updates, forcing them to align along a common axis—the mean user direction, which becomes the popularity direction $\popdir$. The magnitude of this update, as shown in Equation~\ref{eq:item_grad_intuition}, is directly proportional to the item's popularity, $Pop(i)$. This process systematically embeds an item's global popularity as the principal axis of variation in the latent space.

\subsection{Theoretical Analysis of BPR Gradient Sub-optimality}
The existence of a strong $\popdir$ distorts the optimization landscape for user embeddings. For a training triplet $(u,i,j)$, the BPR gradient with respect to the user embedding $\vu$ is:
\begin{equation}
\nabla_{\vu}\mathcal{L}_{uij} = -\sigma(-\vu^T(\vi - \vj))(\vi - \vj).
\end{equation}
The expected total gradient for user $u$ is taken over all their positive items $i \in \setI_u^+$ and all negative items $j \notin \setI_u^+$. Let $w_{uij} = \sigma(-\vu^T(\vi-\vj))$ denote the scalar coefficient derived from the loss derivative. The expected gradient can be decomposed into contributions from positive and negative samples:
\begin{equation}
\nabla_{\vu}\mathcal{L}_u = \E_{i,j}[w_{uij} \vj] - \E_{i,j}[w_{uij} \vi].
\end{equation}
\begin{itemize}[leftmargin=*]
    \item \textbf{Negative Sample Contribution ($\E[w_{uij} \vj]$)}:
    The expectation is over the vast set of unobserved items $j$. Due to data sparsity, this set approximates the global item population. While individual items are diverse, their idiosyncratic preference features tend to cancel out in the aggregate, leaving the average embedding $\E_j[\vj]$ dominated by the common popularity component. Thus, $\E_j[\vj]$ strongly aligns with $\popdir$. This term provides a consistent push, moving $\vu$ away from the popularity direction. This can be interpreted as a \textit{popularity calibration} signal.
    \item \textbf{Positive Sample Contribution ($\E[w_{uij} \vi]$)}: This term is an expectation over the user's interaction history $\setI_u^+$ and should represent the user's unique taste—a \textit{preference signal}. However, if the user has interacted with even a few popular items, their large-magnitude embeddings (aligned with $\popdir$) will disproportionately influence this sum. This "contaminates" the preference signal, pulling the gradient towards the global $\popdir$, even if the user's core taste is for niche items.
\end{itemize}
Ideally, the gradient should point in the Ideal Update Direction, $\dopt \propto (\E_{i \in \setI_u^+}[\vi] - \E_{j \notin \setI_u^+}[\vj])$, which represents the most efficient path to improve ranking. However, due to the popularity contamination in the positive term, the actual BPR gradient is misaligned with $\dopt$. This is the fundamental source of its sub-optimality.

\subsection{The Nature of the Conflict}
The analysis reveals a core conflict: the standard BPR framework forces a single user embedding $\vu$ to perform two distinct and often contradictory tasks simultaneously:
    1) \textbf{Preference Expression}: To rank liked items highly, $\vu$ must have a high dot product with their embeddings $\vi$. This requires $\vu$ to align with the barycenter of the user's true positive items, $\E_{i \in \setI_u^+}[\vi]$.
    2) \textbf{Popularity Calibration}: To rank un-interacted items lowly, $\vu$ must minimize dot products with their embeddings $\vj$. Since the centroid of un-interacted items aligns with the popularity direction (as their specific features largely cancel out), this objective effectively requires $\vu$ to be orthogonal to or pushed away from $\popdir$.

When a user's true preference is for niche items, their ideal preference direction is not aligned with $\popdir$. BPR's confounded gradient forces the update for $\vu$ into a compromised direction—a mixture of the true preference signal and the global popularity signal. This misaligned update is inefficient. It may increase the score for a liked niche item but simultaneously and undesirably increase the scores for all popular items. This dynamic traps the user embedding in a suboptimal local minimum where the gradient may be small, but the representation fails to capture the user's specific tastes accurately.

\subsection{Theoretical Basis for Decoupling}

The mathematical nature of the problem points to a solution: decoupling the two tasks. The standard BPR loss for a triplet is:
\begin{equation}
\mathcal{L}_{uij} = -\ln \sigma(\vu^T \vi - \vu^T \vj).
\end{equation}
Crucially, the same vector $\vu$ is used in both terms of the score difference. We can re-frame this by assigning distinct roles to the user embedding for positive and negative interactions. Consider an equivalent formulation where we use two separate (but initially identical) user vectors:
\begin{equation}
\mathcal{L}_{uij} = -\ln \sigma(({\vu^{pos-term}})^T \vi - {(\vu^{neg-term}})^T \vj),
\end{equation}
where $\vu^{pos-term} = \vu^{eg-term} = \vu$.
The core flaw of BPR is not the loss function itself, but the constraint that $\vu^{pos-term}$ must equal $\vu^{neg-term}$. By relaxing this constraint, we can assign each vector to its specialized task:
\begin{itemize}[leftmargin=*]
    \item The update for $\vu^{pos-term}$ can be restricted to a direction that purely captures user preference.
    \item The update for $\vu^{neg-term}$ can be restricted to a direction that handles popularity calibration.
\end{itemize}
This theoretical reframing suggests that the optimization of a user embedding can be decomposed into updates along two distinct directions: one for preference and one for popularity. This provides the direct theoretical justification for the asymmetric update rule we introduce in the following section.

\begin{figure}[t]
\centering
\includegraphics[width=1.0\linewidth]{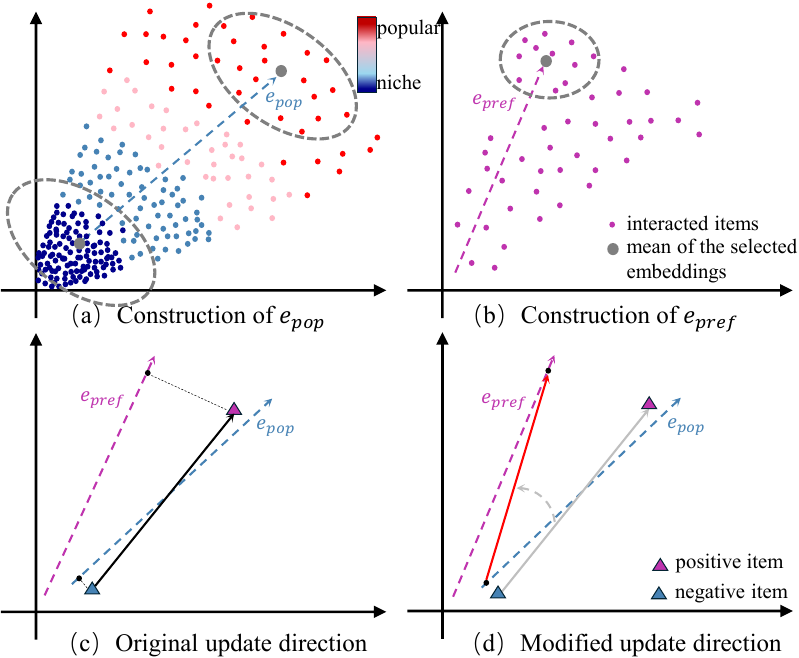}
\caption{Conceptual illustration of our proposed Directional Decomposition and Correction (DDC) framework. (a) We construct the global \textbf{popularity direction ($\popdir$)} by computing the difference vector between the mean embedding of high-popularity items and that of low-popularity items. (b) For each user, we construct a personalized \textbf{preference direction ($\predir$)} from the average embedding of their most preferred items based on their interaction history. (c) The original BPR update direction is a mixture of preference and popularity signals. (d) DDC modifies the update by decomposing it along the $\predir$ and $\popdir$ axes, correcting the gradient to better align with the user's true preference while calibrating for popularity.}
\label{fig:main_observation}
\end{figure}

\section{Directional Decomposition and Correction}
\label{sec:method}
Based on our analysis, we propose a novel fine-tuning framework, \textbf{D}irectional \textbf{D}ecomposition and \textbf{C}orrection (\textbf{DDC}), to rectify the distorted embedding geometry, as illustrated in Figure~\ref{fig:main_observation}. DDC directly implements the decoupling principle derived in Section~\ref{sec:analysis}. Instead of updating the monolithic user embedding $\vu$, we decompose its correction into two targeted, one-dimensional updates along a personalized preference axis and the global popularity axis.

\subsection{Decoupling the BPR Update}
We resolve the gradient conflict by modifying the user representation within the BPR loss using an asymmetric rule. This is performed in a fine-tuning stage where the original model's user embeddings $\vuorig$ and item embeddings $\vi^*$ are frozen. For each user $u$, we learn two scalar coefficients, $\alpha_u$ and $\beta_u$, that control corrections along two pre-defined directions.

 \textbf{Positive Interaction: Preference Alignment.} For a positive pair $(u, i)$, the goal is to reinforce user $u$'s specific taste. The update should move the user embedding towards items they genuinely like. We define a personalized preference direction ($\predir$) for user $u$. A robust way to construct $\predir$ is by leveraging the user's ground-truth interaction history. 
Specifically, we first evaluate the items in user $u$'s history $\setI_u^+$ using the pre-trained model scores $\hat{y}_{ui} = \vuorig^T \vi^*$. Let $\setS_u^{top} \subset \setI_u^+$ be the subset of items constituting the top-$k$ fraction of these scores. We then construct the preference direction by averaging the embeddings of these reliable items:
\begin{equation}
\label{eq:pref_dir_def}
{\predir}_u \defeq \frac{\sum_{i \in \setS_u^{top}} \vi^*}{\norm{\sum_{i \in \setS_u^{top}} \vi^*}}.
\end{equation}
This approach grounds the preference direction in demonstrated user behavior, rather than potentially biased model recommendations.
    While the initial ranking of interacted items might still be partially influenced by the pre-trained model's bias, we treat the proportion $k$ as a key hyperparameter. As shown in our analysis in Section~\ref{sec:Sensitivity_to_Preference}, this allows us to effectively navigate the trade-off between a highly specific (low $k$) and a more robust but potentially popularity-contaminated (high $k$) preference signal. 
    While more sophisticated constructions for $\predir$ are possible, we find this straightforward definition already highly effective.
    For the positive term in the BPR loss, we use a modified user embedding that can only be adjusted along this direction:
    \begin{equation}
    \vu^{pos-term} = \vuorig + \beta_u {\predir}_u,
    \end{equation}
    where $\beta_u$ is a learnable, user-specific scalar that controls the magnitude of movement along the preference direction.
    
     \textbf{Negative Interaction: Popularity Calibration.} For a negative pair $(u, j)$, the objective is to correctly rank an un-interacted item. As established in our analysis, this primarily requires calibrating against global popularity. We therefore use the global popularity direction ($\popdir$) for this task. The effective user embedding for the negative term is:
    \begin{equation}
    \vu^{neg-term} = \vuorig + \alpha_u \popdir,
    \end{equation}
    where $\alpha_u$ is a learnable scalar for popularity calibration. 
    Through optimization, $\alpha_u$ becomes negative, effectively calibrating the user's score profile away from popularity-biased scoring patterns.
\subsection{DDC Loss Function}
By substituting these two asymmetric user representations into the BPR loss, we formulate the DDC fine-tuning objective. For each user $u$, we learn the optimal scalar coefficients $(\alpha_u, \beta_u)$ by minimizing:
\begin{equation}
\label{eq:ddc_loss}
\mbox{\footnotesize $\displaystyle
\begin{aligned}
\mathcal{L}_{DDC} &= \sum_{(u,i,j) \in \mathcal{D}} -\ln \sigma\left( (\vu^{pos-term})^T \vi^* - (\vu^{neg-term})^T \vj^* \right) \\
&= \sum_{(u,i,j) \in \mathcal{D}} -\ln \sigma\left( (\vuorig + \beta_u {\predir}_u)^T \vi^* - (\vuorig + \alpha_u \popdir)^T \vj^* \right)
\end{aligned}
$}
\end{equation}
where $\vi^*$ and $\vj^*$ are the frozen item embeddings from the pre-trained model. This objective disentangles the learning process. The gradient with respect to $\beta_u$ only depends on aligning with positive items, while the gradient with respect to $\alpha_u$ primarily depends on calibrating against negative items.

After fine-tuning and learning the optimal scalars $\alpha_u^*$ and $\beta_u^*$, the final, corrected user embedding for recommendation is constructed by applying both learned corrections:
\begin{equation}
\label{eq:final_embedding}
\vu^{final} \defeq \vuorig + \alpha_u^* \popdir + \beta_u^* {\predir}_u.
\end{equation}
This framework does not increase the dimensionality of the base model. Instead, it provides a principled, low-dimensional correction that guides the user embeddings out of the suboptimal minima created by standard BPR, leading to an accelerated optimization dynamic in the efficient fine-tuning stage and significantly improved recommendation performance.

\section{Experiments}
In this section, we conduct extensive experiments to answer the following research questions:
\begin{itemize}[leftmargin=*]
\item  \textbf{RQ1}: How effectively does our DDC framework enhance recommendation accuracy, and does it achieve this by successfully mitigating popularity bias?
\item \textbf{RQ2}: How do the different components of DDC, particularly the asymmetric update rule and the final embedding composition, contribute to its effectiveness?
\item \textbf{RQ3}: How sensitive is DDC to its key hyperparameters, and how does it impact model convergence?
\end{itemize}
\subsection{Experimental Setup}

\subsubsection{Datasets}
We evaluate our method on three widely-used public benchmark datasets with varying characteristics and sparsity. To ensure the quality of the dataset, we use the 10-core setting. The statistics are summarized in Table~\ref{tab:datasets}.

\begin{table}[t]
\centering
\caption{Statistics of the experimental datasets.}
\label{tab:datasets}
\resizebox{1.0\columnwidth}{!}{%
\setlength{\aboverulesep}{0pt}
\setlength{\belowrulesep}{0.45pt} 
\begin{tabular}{lrrrr}
\toprule
\textbf{Dataset} & \textbf{\#Users} & \textbf{\#Items} & \textbf{\#Interactions} & \textbf{Sparsity} \\
\midrule
Amazon-Book & 139,090 & 113,176 & 3,344,074 & 99.979\% \\
Yelp & 135,868 & 68,825 & 3,857,030 & 99.959\% \\
Tmall & 125,554 & 58,059 & 2,064,290 & 99.972\% \\
\bottomrule
\end{tabular}%
}
\end{table}

\subsubsection{Baselines}
We compare DDC against two groups of baselines:
\begin{itemize}[leftmargin=1em]
    \item \textbf{Backbone Models}: We show the general applicability of DDC by applying it to five representative models: 
    MF~\cite{BPRRendle2009}, LightGCN~\cite{LightGCNHe2020}, DGCF~\cite{DisentangledGraphWang2020}, NCL~\cite{ImprovingGraphLin2022}, and LightCCF~\cite{UnveilingContrastiveZhang2025}.
    \item \textbf{Debiasing Methods}: We compare DDC with seven state-of-the-art debiasing methods: IPS~\cite{UnbiasedLearningJoachims2017}, DICE~\cite{DisentanglingUserZheng2021}, MACR~\cite{ModelAgnosticWei2021}, PC~\cite{PopularityOpportunityZhu2021}, PAAC~\cite{PopularityAwareCai2024}, DCCL~\cite{DCCL}, and TPAB~\cite{TPAB}. For a fair comparison, these methods are applied to two backbone models, MF and LightGCN.
\end{itemize}
Our method, DDC, is applied as a fine-tuning stage to pre-trained backbones (e.g., LightGCN-DDC).
\subsubsection{Evaluation Metrics}
We evaluate the top-N recommendation performance using three standard metrics: Recall@10, NDCG@10, and MRR@10. Higher values indicate better performance.
\subsubsection{Implementation Details}
To ensure a fair comparison, we set the training batch size to 8192 and implement all methods under the same framework, RecBole~\cite{recbole[1.0],recbole[2.0],recbole[1.2.1]}. For all models, including backbones and debiasing baselines, we performed a grid search to find optimal hyperparameters. Crucially, we did not set an upper limit on epochs; all models were trained until convergence, determined by an early stopping strategy where training terminates if validation performance (MRR@10) does not improve for 50 consecutive epochs. The model achieving the best validation performance was then evaluated on the test set. The embedding dimension is set to $d=64$. For our DDC method, the only significant hyperparameter is the proportion of a user's interacted items used to construct their personalized preference direction $\predir$. While careful tuning of this proportion can yield further gains, for the main experiments in Tables~\ref{tab:backbone_results} and~\ref{tab:debiasing_comparison}, we uniformly set this value to 30\% to demonstrate robust performance without extensive hyperparameter tuning.
\subsection{Performance and Bias Analysis (RQ1)}
\subsubsection{Effectiveness on Various Backbone Models}
To answer RQ1, we first evaluate DDC's ability to serve as a plug-and-play module for enhancing various converged recommendation models. Table~\ref{tab:backbone_results} presents the performance of five backbones before and after applying our DDC fine-tuning.

\begin{table*}[t]
\centering
\caption{Performance improvement after applying DDC to five different backbone models. DDC consistently and significantly boosts the performance of all backbones across all datasets, demonstrating its model-agnostic effectiveness. The values in \textbf{bold} denote the results of the DDC-enhanced models, with the corresponding relative improvements shown below.}
\label{tab:backbone_results}
\renewcommand{\arraystretch}{0.85} 
\resizebox{\textwidth}{!}{%
\begin{tabular}{l|ccc|ccc|ccc}
\toprule
& \multicolumn{3}{c|}{\textbf{Amazon-Book}} & \multicolumn{3}{c|}{\textbf{Yelp}} & \multicolumn{3}{c}{\textbf{Tmall}} \\
\textbf{Method} & \textbf{MRR@10} & \textbf{NDCG@10} & \textbf{MAP@10} & \textbf{MRR@10} & \textbf{NDCG@10} & \textbf{MAP@10} & \textbf{MRR@10} & \textbf{NDCG@10} & \textbf{MAP@10} \\
\midrule
MF & 0.0557 & 0.0444 & 0.0272 & 0.0588 & 0.0410 & 0.0236 & 0.0599 & 0.0490 & 0.0323 \\
\textbf{MF-DDC} & \textbf{0.0660} & \textbf{0.0520} & \textbf{0.0325} & \textbf{0.0760} & \textbf{0.0502} & \textbf{0.0308} & \textbf{0.0677} & \textbf{0.0552} & \textbf{0.0366} \\
\rowcolor{lightgray} 
Improvement & +18.5\% & +17.1\% & +19.5\% & +29.3\% & +22.4\% & +30.5\% & +13.0\% & +12.7\% & +13.3\% \\
\midrule
LightGCN & 0.0709 & 0.0563 & 0.0354 & 0.0766 & 0.0534 & 0.0320 & 0.0670 & 0.0558 & 0.0366 \\
\textbf{LightGCN-DDC} & \textbf{0.0814} & \textbf{0.0640} & \textbf{0.0406} & \textbf{0.0860} & \textbf{0.0578} & \textbf{0.0354} & \textbf{0.0737} & \textbf{0.0605} & \textbf{0.0402} \\
\rowcolor{lightgray} 
Improvement & +14.8\% & +13.7\% & +14.7\% & +12.3\% & +8.2\% & +10.6\% & +10.0\% & +8.4\% & +9.8\% \\
\midrule
DGCF & 0.0603 & 0.0476 & 0.0294 & 0.0683 & 0.0479 & 0.0281 & 0.0612 & 0.0501 & 0.0330 \\
\textbf{DGCF-DDC} & \textbf{0.0715} & \textbf{0.0559} & \textbf{0.0352} & \textbf{0.0782} & \textbf{0.0528} & \textbf{0.0320} & \textbf{0.0693} & \textbf{0.0565} & \textbf{0.0376} \\
\rowcolor{lightgray} 
Improvement & +18.6\% & +17.4\% & +19.7\% & +14.5\% & +10.2\% & +13.9\% & +13.2\% & +12.8\% & +13.9\% \\
\midrule
NCL & 0.0716 & 0.0567 & 0.0358 & 0.0770 & 0.0533 & 0.0320 & 0.0638 & 0.0525 & 0.0346 \\
\textbf{NCL-DDC} & \textbf{0.0811} & \textbf{0.0635} & \textbf{0.0406} & \textbf{0.0859} & \textbf{0.0579} & \textbf{0.0355} & \textbf{0.0691} & \textbf{0.0564} & \textbf{0.0375} \\
\rowcolor{lightgray} 
Improvement & +13.3\% & +12.0\% & +13.4\% & +11.6\% & +8.6\% & +10.9\% & +8.3\% & +7.4\% & +8.4\% \\
\midrule
LightCCF & 0.0718 & 0.0570 & 0.0357 & 0.0761 & 0.0527 & 0.0312 & 0.0681 & 0.0566 & 0.0372 \\
\textbf{LightCCF-DDC} & \textbf{0.0800} & \textbf{0.0627} & \textbf{0.0397} & \textbf{0.0829} & \textbf{0.0559} & \textbf{0.0338} & \textbf{0.0722} & \textbf{0.0595} & \textbf{0.0393} \\
\rowcolor{lightgray} 
Improvement & +11.4\% & +10.0\% & +11.2\% & +8.9\% & +6.1\% & +8.3\% & +6.0\% & +5.1\% & +5.6\% \\
\bottomrule
\end{tabular}%
}
\end{table*}

The results unequivocally demonstrate that DDC provides substantial and consistent improvements across all five backbone models and three datasets. The improvements are not marginal; for instance, DDC boosts MF's MRR@10 on Yelp by a remarkable 29.3\% and DGCF's on Amazon-Book by 18.6\%. The fact that DDC enhances models ranging from the classic MF to advanced GNN (LightGCN, DGCF) and contrastive learning (NCL, LightCCF) architectures provides powerful evidence for our central claim: the geometric distortion caused by BPR is a fundamental and widespread issue. DDC's ability to rectify this geometry offers a universal and effective solution, unlocking performance potential that was previously trapped in suboptimal local minima.

\subsubsection{Comparison with State-of-the-Art Debiasing Methods}
Next, we compare DDC with seven competitive debiasing baselines, using MF and LightGCN as base models. The results are detailed in Table~\ref{tab:debiasing_comparison}.

\begin{table*}[t]
\centering
\caption{Comparison with SOTA debiasing methods using MF and LightGCN as backbones. The best performance for each backbone is in \textbf{bold}, and the second-best is \underline{underlined}. DDC consistently outperforms all specialized debiasing methods on both backbones and all datasets.}
\label{tab:debiasing_comparison}
\renewcommand{\arraystretch}{0.85} 
\resizebox{\textwidth}{!}{%
\begin{tabular}{l|ccc|ccc|ccc}
\toprule
& \multicolumn{3}{c|}{\textbf{Amazon-Book}} & \multicolumn{3}{c|}{\textbf{Yelp}} & \multicolumn{3}{c}{\textbf{Tmall}} \\
\textbf{Method} & \textbf{MRR@10} & \textbf{NDCG@10} & \textbf{MAP@10} & \textbf{MRR@10} & \textbf{NDCG@10} & \textbf{MAP@10} & \textbf{MRR@10} & \textbf{NDCG@10} & \textbf{MAP@10} \\
\midrule
MF & 0.0557 & 0.0444 & 0.0272 & \underline{0.0588} & \underline{0.0410} & \underline{0.0236} & 0.0599 & \underline{0.0490} & \underline{0.0323} \\
MF-IPS & 0.0358 & 0.0294 & 0.0186 & 0.0283 & 0.0194 & 0.0105 & 0.0413 & 0.0300 & 0.0214 \\
MF-DICE & 0.0492 & 0.0386 & 0.0235 & 0.0510 & 0.0345 & 0.0192 & 0.0586 & 0.0481 & 0.0316 \\
MF-MACR & 0.0505 & 0.0405 & 0.0248 & 0.0451 & 0.0313 & 0.0172 & 0.0563 & 0.0457 & 0.0301 \\
MF-PC & 0.0299 & 0.0243 & 0.0149 & 0.0178 & 0.0123 & 0.0063 & 0.0411 & 0.0298 & 0.0213 \\
MF-PAAC & 0.0557 & 0.0443 & 0.0273 & 0.0577 & 0.0398 & 0.0228 & 0.0593 & 0.0484 & 0.0318 \\
MF-DCCL & 0.0564 & 0.0445 & 0.0274 & 0.0585 & 0.0406 & 0.0233 & 0.0594 & 0.0485 & 0.0319 \\
MF-TPAB & \underline{0.0565} & \underline{0.0450} & \underline{0.0276} & 0.0580 & 0.0406 & 0.0232 & \underline{0.0602} & \underline{0.0490} & 0.0322 \\
\textbf{MF-DDC} & \textbf{0.0660} & \textbf{0.0520} & \textbf{0.0325} & \textbf{0.0760} & \textbf{0.0502} & \textbf{0.0308} & \textbf{0.0677} & \textbf{0.0552} & \textbf{0.0366} \\
\rowcolor{lightgray} 
Improvement & +16.8\% & +15.6\% & +17.8\% & +29.3\% & +22.4\% & +30.5\% & +12.5\% & +12.7\% & +13.3\% \\
\midrule
\midrule
LightGCN & 0.0709 & 0.0563 & 0.0354 & 0.0766 & 0.0534 & 0.0320 & 0.0670 & 0.0558 & 0.0366 \\
LightGCN-IPS & 0.0348 & 0.0286 & 0.0170 & 0.0269 & 0.0178 & 0.0093 & 0.0367 & 0.0317 & 0.0201 \\
LightGCN-DICE & 0.0664 & 0.0524 & 0.0328 & 0.0770 & 0.0528 & 0.0318 & 0.0643 & 0.0543 & 0.0351 \\
LightGCN-MACR & 0.0293 & 0.0239 & 0.0142 & 0.0365 & 0.0250 & 0.0138 & 0.0528 & 0.0438 & 0.0284 \\
LightGCN-PC & 0.0713 & 0.0567 & 0.0357 & 0.0764 & 0.0532 & 0.0317 & 0.0667 & 0.0556 & 0.0366 \\
LightGCN-PAAC & \underline{0.0794} & \underline{0.0630} & \underline{0.0394} & 0.0781 & 0.0534 & 0.0307 & \underline{0.0707} & \underline{0.0592} & \underline{0.0383} \\
LightGCN-DCCL & 0.0728 & 0.0578 & 0.0364 & 0.0772 & 0.0535 & 0.0319 & 0.0682 & 0.0565 & 0.0371 \\
LightGCN-TPAB & 0.0777 & 0.0615 & 0.0392 &  \underline{0.0782} &  \underline{0.0544} &  \underline{0.0323} & 0.0674 & 0.0560 & 0.0367 \\
\textbf{LightGCN-DDC} & \textbf{0.0814} & \textbf{0.0640} & \textbf{0.0406} & \textbf{0.0860} & \textbf{0.0578} & \textbf{0.0354} & \textbf{0.0737} & \textbf{0.0605} & \textbf{0.0402} \\
\rowcolor{lightgray} 
Improvement & +2.5\% & +1.6\% & +3.0\% & +10.0\% & +6.3\% & +9.6\% & +4.2\% & +2.2\% & +5.0\% \\
\bottomrule
\end{tabular}%
}
\end{table*}

The results are compelling. DDC decisively outperforms all other debiasing methods. Notably, many existing methods like IPS and MACR even degrade performance, suggesting that macroscopic approaches like re-weighting or complex causal modeling can be unstable or rely on faulty assumptions. Other methods like PAAC and TPAB provide some gains but are far surpassed by DDC. For example, on Yelp with LightGCN, LightGCN-DDC achieves an MRR@10 of 0.0860, a 10.0\% relative improvement over the strongest baseline, LightGCN-TPAB. This superior performance strongly suggests that our method of directly identifying and correcting the geometric source of popularity bias is a more fundamental and effective solution than treating its symptoms.
\subsubsection{Analysis of Popularity Bias Mitigation}
A core claim of our paper is that DDC improves recommendation accuracy by mitigating popularity bias at its geometric source. To directly validate this, we examine how DDC impacts the popularity of recommended items. We use the AvgPop@10 metric, which calculates the average interaction count of the top-10 recommended items across all test users. The formal definition is:
$$ \mathrm{AvgPop@K} = \frac{1}{|U|} \sum_{u \in U } \frac{\sum_{i \in R_{u}} \phi(i)}{|R_{u}|}, $$
where $R_u$ is the set of top-K recommended items for user $u$, and $\phi(i)$ is the total number of interactions for item $i$ in the training data. A lower value indicates the model is recommending less popular, more diverse items. We present the results for all five backbone models on the Tmall dataset in Table~\ref{tab:popularity_analysis}.

\begin{table}[h]
\centering
\caption{Impact of DDC on recommendation accuracy and popularity on the Tmall dataset. DDC simultaneously improves accuracy metrics while significantly reducing the average popularity of recommended items, confirming its effectiveness in mitigating popularity bias.}
\label{tab:popularity_analysis}

\renewcommand{\arraystretch}{0.9} 

\resizebox{\columnwidth}{!}{%

\begin{tabular}{l|cc|r|r}
\toprule
\textbf{Method} & \textbf{MRR@10} & \textbf{NDCG@10} & \textbf{AvgPop@10} $\downarrow$ & \textbf{Change (\%)} \\
\midrule
MF & 0.0599 & 0.0490 & 1472.90 & - \\
\textbf{MF-DDC} & \textbf{0.0677} & \textbf{0.0552} & \textbf{967.18} & \textbf{-34.3\%} \\
\midrule
LightGCN & 0.0670 & 0.0558 & 1642.81 & - \\
\textbf{LightGCN-DDC} & \textbf{0.0737} & \textbf{0.0605} & \textbf{1000.53} & \textbf{-39.1\%} \\
\midrule
DGCF & 0.0612 & 0.0501 & 1563.44 & - \\
\textbf{DGCF-DDC} & \textbf{0.0693} & \textbf{0.0565} & \textbf{997.90} & \textbf{-36.2\%} \\
\midrule
NCL & 0.0638 & 0.0525 & 1248.60 & - \\
\textbf{NCL-DDC} & \textbf{0.0691} & \textbf{0.0564} & \textbf{980.97} & \textbf{-21.4\%} \\
\midrule
LightCCF & 0.0681 & 0.0566 & 1565.37 & - \\
\textbf{LightCCF-DDC} & \textbf{0.0722} & \textbf{0.0595} & \textbf{826.54} & \textbf{-47.2\%} \\
\bottomrule
\end{tabular}%
}
\end{table}

The results in Table~\ref{tab:popularity_analysis} are unequivocal. Across every backbone, DDC not only improves recommendation accuracy but also dramatically reduces the average popularity of recommended items. For example, LightGCN-DDC reduces the AvgPop@10 metric by 39.1\%, and LightCCF-DDC achieves a remarkable 47.2\% reduction, all while significantly boosting MRR@10 and NDCG@10. This provides powerful, direct evidence that our method is working as intended. The concurrent gains in accuracy and the reduction in popularity demonstrate that DDC is not making a simple trade-off between relevance and novelty. Instead, by rectifying the underlying geometric flaw, it allows the model to escape popularity-driven local minima and discover more personalized items, leading to recommendations that are both more accurate and less biased.
\subsection{Ablation and Component Analysis (RQ2)}
To validate the design choices of DDC, we conduct a detailed analysis on the Tmall dataset with MF as the backbone. We analyze two key aspects: the asymmetric update rule and the composition of the final user embedding.

\subsubsection{Effectiveness of the Asymmetric Update Rule}
We test different update strategies within the DDC loss. Let $ \texttt{a} = \alpha_u \popdir$ be the popularity correction and $ \texttt{b} = \beta_u \predir$ be the preference correction. We test nine variants based on which correction is applied to the positive and negative terms in the BPR loss. A rule is denoted as \texttt{`pos-term\_neg-term'}. Our proposed method is \texttt{b\_a}. The results are shown in Table~\ref{tab:ablation}. We can observe that our proposed asymmetric rule, DDC (\texttt{b\_a}), significantly outperforms all other configurations. This result stems from its clear separation of concerns, which resolves the core conflict in standard BPR.
\begin{itemize}[leftmargin=1em]
    \item \textbf{Why \texttt{b\_a} excels}: This rule achieves task separation by directing positive pair updates \textit{only} along personal preference directions ($\predir$) to reinforce individual taste, while restricting negative pair updates to popularity directions ($\popdir$) for global calibration. This disentangled gradient control prevents interference between preference learning and popularity calibration.
    \item \textbf{Why others are suboptimal}: Configurations like \texttt{a\_*} fail because they incorrectly apply the popularity correction to positive items, suppressing the user's preference signal. The symmetric rules \texttt{a\_a} and \texttt{b\_b} perform comparably to or better than the baseline but are still limited because they use a single lens for both preference and popularity. The worst performers, such as \texttt{a\_b}, directly fight against the learning objective. The \texttt{ab\_*} variants, which apply both corrections, re-introduce the confounding effect we aim to eliminate, leading to confused gradients and inferior performance. This confirms the necessity of our asymmetric design.
\end{itemize}

\begin{table}[t]
	\centering
	\caption{Ablation and component analysis on the Tmall dataset using MF. Our proposed asymmetric update rule (DDC (\texttt{b\_a})) and full final embedding are superior.}
	\label{tab:ablation}
    \renewcommand{\arraystretch}{0.9} 
	\resizebox{\columnwidth}{!}{%

	\begin{tabular}{l|ccc}
	\toprule
	\textbf{Variant} & \textbf{MRR@10} & \textbf{NDCG@10} & \textbf{MAP@10} \\
	\midrule
	\multicolumn{4}{l}{\textit{Analysis of Asymmetric Update Rule}} \\
	MF (BPR Baseline) & 0.0599 & 0.0490 & 0.0323 \\
	DDC (\texttt{a\_a}) & 0.0590 & 0.0482 & 0.0317 \\
	DDC (\texttt{a\_b}) & 0.0417 & 0.0323 & 0.0225 \\
	DDC (\texttt{a\_ab}) & 0.0424 & 0.0331 & 0.0229 \\
	DDC (\texttt{b\_b}) & 0.0645 & 0.0528 & 0.0349 \\
	DDC (\texttt{b\_ab}) & 0.0639 & 0.0526 & 0.0345 \\
	DDC (\texttt{ab\_a}) & 0.0674 & 0.0550 & 0.0364 \\
	DDC (\texttt{ab\_b}) & 0.0588 & 0.0476 & 0.0316 \\
	DDC (\texttt{ab\_ab}) & 0.0644 & 0.0527 & 0.0349 \\
	\textbf{DDC (\texttt{b\_a}) (Ours)} & \textbf{0.0677} & \textbf{0.0552} & \textbf{0.0366} \\
	\midrule
	\multicolumn{4}{l}{\textit{Analysis of Final Embedding Composition}} \\
	DDC (w/o $\alpha_u^* \popdir$) & 0.0672 & 0.0548 & 0.0363 \\
	DDC (w/o $\beta_u^* \predir$) & 0.0591 & 0.0486 & 0.0318 \\
	\textbf{DDC (full, Eq.~\ref{eq:final_embedding})} & \textbf{0.0677} & \textbf{0.0552} & \textbf{0.0366} \\
	\bottomrule
	\end{tabular}%
	}
\end{table}

\subsubsection{Effectiveness of Final Embedding Composition}
We test the contribution of each directional component in the final user embedding (Eq.~\ref{eq:final_embedding}). As shown in the bottom half of Table~\ref{tab:ablation}, removing either correction vector from the final embedding composition leads to a drop in performance.
Removing the preference alignment ($\beta_u^* \predir$) is particularly detrimental, causing performance to collapse nearly to the baseline level. This confirms that enhancing the true preference signal is the primary driver of improvement. However, removing the popularity correction ($\alpha_u^* \popdir$) also results in a noticeable performance decrease, indicating that explicit calibration against global popularity is vital for achieving the best results. This validates our design of combining both learned corrections to form the final, rectified user embedding.

\subsection{Parameter Sensitivity and Convergence Analysis (RQ3)}
\subsubsection{Sensitivity to Preference Direction Granularity}
\label{sec:Sensitivity_to_Preference}
\begin{figure}[t]
    \centering
    \includegraphics[width=\columnwidth]{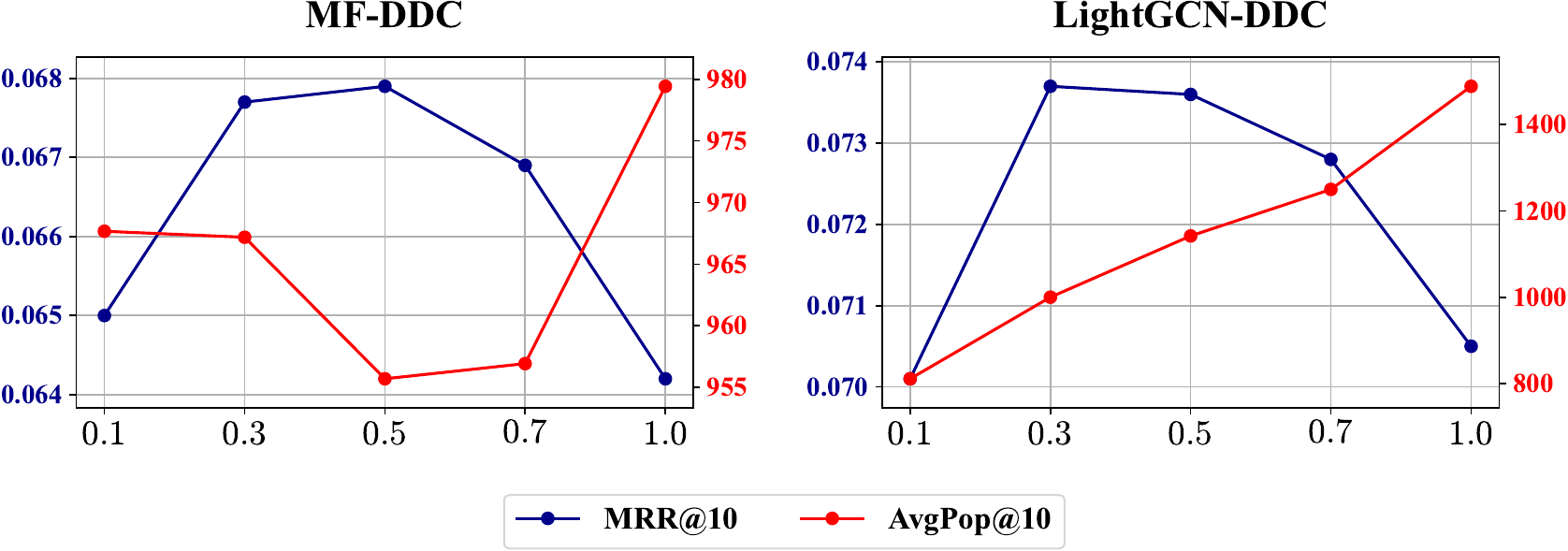} 
    \caption{Parameter sensitivity analysis of the proportion $k$ on the Tmall dataset, showing its dual impact on recommendation accuracy.
    }
    \label{fig:param_k}
\end{figure}

We investigate the sensitivity of DDC to its key hyperparameter, $k$, which defines the proportion of a user's most relevant interacted items used to construct their personalized preference direction $\predir$. We evaluate its impact on both recommendation accuracy (MRR@10) and popularity bias (AvgPop@10) for MF and LightGCN backbones on the Tmall dataset. The results are shown in Figure~\ref{fig:param_k}.

The analysis reveals a critical trade-off between accuracy and bias mitigation. For both MF-DDC and LightGCN-DDC, the recommendation accuracy (MRR@10) follows a concave trend, peaking at intermediate values of $k$ ($k=0.5$ for MF and $k=0.3$ for LightGCN). This suggests that:
\begin{itemize}[leftmargin=1em]
    \item A small $k$ (e.g., 0.1) creates a preference direction $\predir$ from too few items, making the signal potentially noisy and failing to capture the user's full interest profile, thus slightly hurting accuracy.
    \item A large $k$ (e.g., 1.0) includes less relevant items, making the preference direction too generic and pulling it closer to the global popularity distribution, which also degrades personalized recommendation accuracy.
\end{itemize}

The AvgPop@10 metric provides deeper insight into the effect of $k$. For LightGCN-DDC, we observe a strong positive correlation between $k$ and the average popularity of recommended items. As $k$ increases, the model recommends significantly more popular items, confirming that a larger $k$ dilutes the personalized signal with global popularity. For MF-DDC, the trend is more subtle, but the lowest popularity bias is achieved at $k=0.5$, which gratifyingly aligns with its peak accuracy.

In conclusion, the choice of $k$ navigates a trade-off between a highly specific, low-popularity-bias signal (low $k$) and a more robust but generic, high-popularity-bias signal (high $k$). The results show that DDC is robust across a reasonable range of $k$. Our choice of \textbf{$k=0.3$} for the main experiments is a well-justified one, as it achieves near-optimal accuracy for both backbones while effectively keeping popularity bias in check, especially for the more powerful LightGCN model.

\subsubsection{Convergence Analysis}~\label{sec:conv}
\begin{figure}[t]
    \centering
    \includegraphics[width=0.48\textwidth]{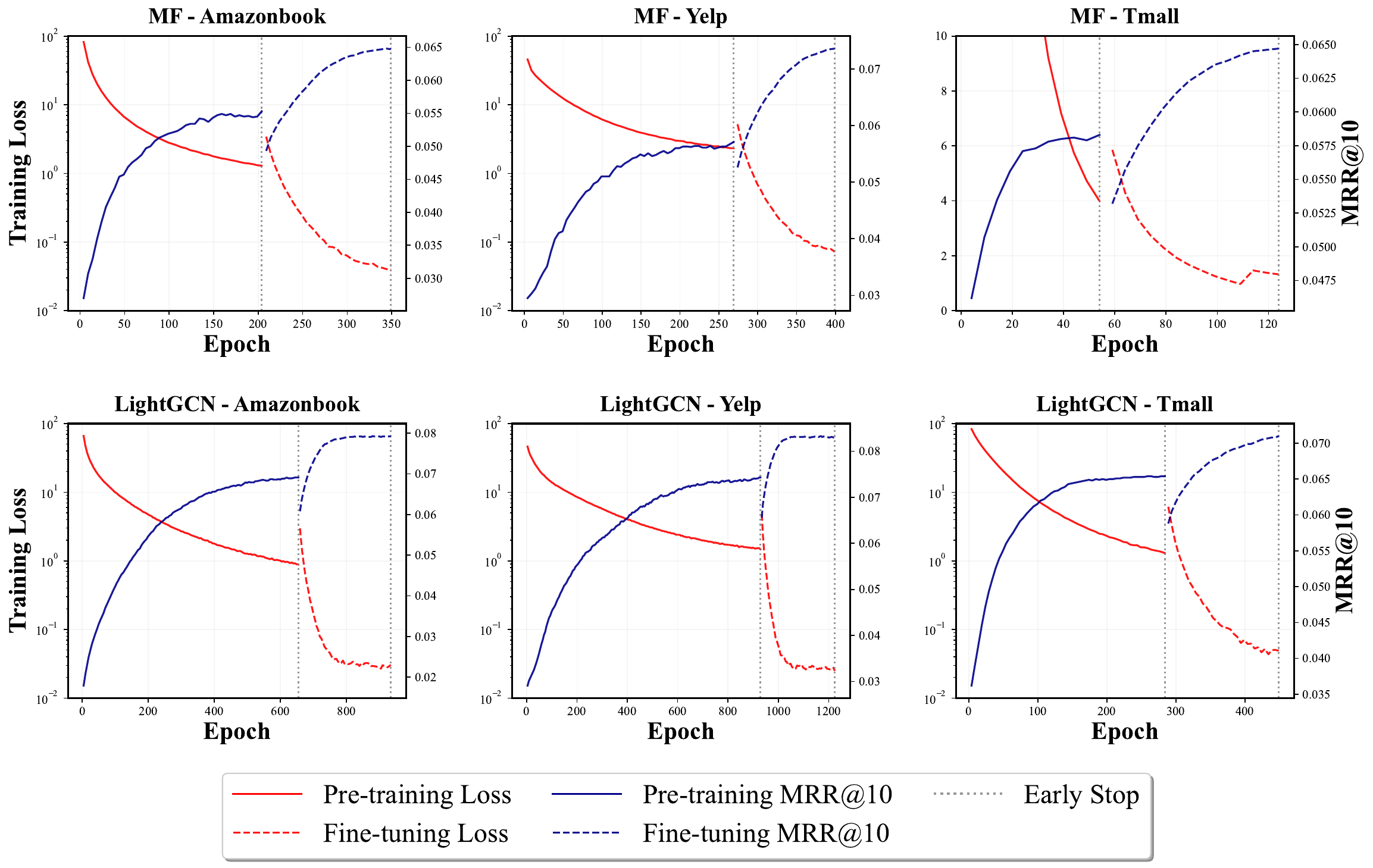} 
    \caption{Convergence curves of BPR loss and MRR@10 for MF and LightGCN on three datasets. 
    }
    \label{fig:convergence}
\end{figure}

Finally, we analyze DDC's impact on convergence. The left side of Figure~\ref{fig:convergence} shows that standard BPR training causes the loss to slowly decrease and plateau at a high value. However, once the backbone model converges and DDC fine-tuning begins, the BPR loss plummets dramatically. Note that because the DDC correction coefficients, $\alpha_u$ and $\beta_u$, are randomly initialized, the values at the start of the fine-tuning phase do not perfectly align with the backbone's converged values, causing an initial jump. To ensure a fair comparison, the loss shown during the DDC phase is not its optimization objective (Eq.~\ref{eq:ddc_loss}), but rather the original BPR loss calculated with the final corrected embeddings:
$$ \mathcal{L}_{eval} = \sum_{(u,i,j) \in \mathcal{D}} -\ln \sigma\left( (\vu^{final})^T \vi^* - (\vu^{final})^T \vj^* \right). $$

Quantitatively, on the Yelp dataset, the LightGCN baseline converges after 929 epochs with a final loss of 1.5055. After DDC fine-tuning, the loss drops to just 0.0267 (about 1.8\% of the original). Similarly, for MF on Amazon-Book, the loss drops from 1.2922 to 0.0191 (about 1.5\% of the original). This reflects a highly efficient optimization trajectory where the model rapidly breaks through previous performance ceilings to reach a fundamentally superior solution. As seen on the right side of Figure~\ref{fig:convergence}, this massive reduction in loss corresponds to a rapid and substantial climb in MRR@10, quickly surpassing the baseline's peak performance. This provides powerful, direct evidence for our thesis: DDC's principled correction allows embeddings to escape the suboptimal local minima created by BPR's geometric bias, finding a much better representation of true user preference.

\section{Related Work}
The related works can be grouped into three categories:

\noindent\textbf{Macroscopic Debiasing Strategies.}
Early approaches to mitigating popularity bias treat reommendation models as black boxes, intervening at the data or objective function level. For example, reweighting methods, like inversion propensity scoring (IPS)~\cite{CounterfactualReasoningBottou2013,AutoDebiasChen2021,PopularityOpportunityZhu2021}, down-weight popular item interactions to amplify signals from niche items, but can suffer from high variance and oversimplification~\cite{UnbiasedLearningJoachims2017,RecommendationsasTreatmentsSchnabel2016}. Regularization techniques add penalties to the objective function to discourage correlation with item popularity, often at the cost of overall recommendation accuracy~\cite{IncorporatingDiversityJacek2016,ControllingPopularityAbdollahpouri2017,PopularityOpportunityZhu2021,10.1145/3711896.3737068}. Critically, these macroscopic strategies address the symptoms of popularity bias without dissecting its underlying structural impact on the model's learned representations.

\noindent\textbf{Causal and Disentangled Approaches.}
More principled solutions leverage causal inference and representation disentanglement. Causal approaches use graphs to model and remove popularity's confounding influence, often via counterfactual inference~\cite{DeconfoundedRecommendationWang2021,CausalInterventionZhang2021,CausalEmbeddingsBonner2018}. Similarly, disentanglement methods aim to learn separate latent factors for genuine interest versus conformity to popular trends~\cite{DisentanglingUserZheng2021}, with the goal of isolating a  `pure' preference representation for prediction. While sophisticated, these methods often rely on strong causal assumptions or complex training schemes~\cite{ModelAgnosticWei2021}. Crucially, they hypothesize a popularity component but fail to explain how it is structurally and geometrically encoded by standard optimization like BPR—a foundational gap our work directly addresses.

\noindent\textbf{Representation Space Optimization.}
Contrastive learning (CL) has emerged as a powerful paradigm to improve embedding quality in GNN-based models~\cite{SelfSupervisedWu2021,AreGraphAugmentationsYu2022}. CL-based methods like SGL~\cite{SelfSupervisedWu2021} and SimGCL~\cite{AreGraphAugmentationsYu2022} employ a contrastive loss to foster a more uniform distribution of embeddings, which can indirectly alleviate popularity bias by preventing popular items from dominating the latent space~\cite{UnderstandingContrastiveWang2020,TowardsRepresentationWang2022}. However, these approaches pursue uniformity as a general objective for representation quality; they do not directly address the specific, systematic geometric distortion that popularity imprints on the embedding space. Therefore, while beneficial, this pursuit of general uniformity is an indirect and incomplete solution to the specific geometric problem of popularity bias. Our work differs by identifying the principal ``popularity direction'' and then applying a targeted, asymmetric correction, fundamentally realigning the space to disentangle preference from popularity.
\section{Conclusion}

This work reveals that popularity bias in CF stems from a fundamental geometric distortion within BPR optimization itself. We identify a dominant popularity direction that systematically conflates item quality with interaction frequency, explaining why traditional debiasing methods achieve limited success—they address symptoms while leaving the underlying geometric bias intact.
Our {D}irectional {D}ecomposition and {C}orrection framework directly corrects this distortion through asymmetric directional updates. DDC decomposes gradients into popularity and preference components, reinforcing personal taste along preference directions while calibrating against popularity bias along popularity direction. DDC's universality across BPR architectures validates that addressing geometric bias fundamentally improves recommendation quality.

\begin{acks}
This work was supported by the New Generation Artificial Intelligen\allowbreak ce-National Science and Technology Major Project (Grant No. 2025ZD01\allowbreak 22601), the National Natural Science Foundation of China (Grant Nos. 62506348 and 62406141), the Natural Science Foundation of Anhui Province (Grant No. 2508085QF211), the CCF-1688 Yuanbao Cooperation Fund (Grant No. CCF-Alibaba2025005), the China Postdoctoral Science Foundation (Grant No. GZC20252740), and the Opening Foundation of State Key Laboratory of Cognitive Intelligence, iFLYTEK (Grant No. COGOS-2025HE02).
\end{acks}

\balance
\bibliographystyle{ACM-Reference-Format}
\bibliography{References}

@inproceedings{IncorporatingDiversityJacek2016,
  title = {Incorporating Diversity in a Learning to Rank Recommender System},
  author = {Jacek Wasilewski and Neil Hurley},
  year = {2016},
  url = {http://www.aaai.org/ocs/index.php/FLAIRS/FLAIRS16/paper/view/12944},
  researchr = {https://researchr.org/publication/WasilewskiH16},
  cites = {0},
  citedby = {0},
  pages = {572-578},
  booktitle = {Proceedings of the Twenty-Ninth International Florida Artificial Intelligence Research Society Conference, FLAIRS 2016, Key Largo, Florida, May 16-18, 2016},
  editor = {Zdravko Markov and Ingrid Russell},
  publisher = {AAAI Press},
  isbn = {978-1-57735-756-8},
}

@article{CounterfactualReasoningBottou2013,
author = {Bottou, L\'{e}on and Peters, Jonas and Qui\~{n}onero-Candela, Joaquin and Charles, Denis X. and Chickering, D. Max and Portugaly, Elon and Ray, Dipankar and Simard, Patrice and Snelson, Ed},
title = {Counterfactual reasoning and learning systems: the example of computational advertising},
year = {2013},
issue_date = {January 2013},
publisher = {JMLR.org},
volume = {14},
number = {1},
issn = {1532-4435},
journal = {J. Mach. Learn. Res.},
month = jan,
pages = {3207–3260},
numpages = {54}
}

@inproceedings{AutoDebiasChen2021,
author = {Chen, Jiawei and Dong, Hande and Qiu, Yang and He, Xiangnan and Xin, Xin and Chen, Liang and Lin, Guli and Yang, Keping},
title = {AutoDebias: Learning to Debias for Recommendation},
year = {2021},
isbn = {9781450380379},
publisher = {Association for Computing Machinery},
address = {New York, NY, USA},
url = {https://doi.org/10.1145/3404835.3462919},
doi = {10.1145/3404835.3462919},
booktitle = {Proceedings of the 44th International ACM SIGIR Conference on Research and Development in Information Retrieval},
pages = {21–30},
numpages = {10},
location = {Virtual Event, Canada},
series = {SIGIR '21}
}

@inproceedings{PopularityOpportunityZhu2021,
author = {Zhu, Ziwei and He, Yun and Zhao, Xing and Zhang, Yin and Wang, Jianling and Caverlee, James},
title = {Popularity-Opportunity Bias in Collaborative Filtering},
year = {2021},
isbn = {9781450382977},
publisher = {Association for Computing Machinery},
address = {New York, NY, USA},
url = {https://doi.org/10.1145/3437963.3441820},
doi = {10.1145/3437963.3441820},
booktitle = {Proceedings of the 14th ACM International Conference on Web Search and Data Mining},
pages = {85–93},
numpages = {9},
location = {Virtual Event, Israel},
series = {WSDM '21}
}

@inproceedings{UnbiasedLearningJoachims2017,
author = {Joachims, Thorsten and Swaminathan, Adith and Schnabel, Tobias},
title = {Unbiased Learning-to-Rank with Biased Feedback},
year = {2017},
isbn = {9781450346757},
publisher = {Association for Computing Machinery},
address = {New York, NY, USA},
url = {https://doi.org/10.1145/3018661.3018699},
doi = {10.1145/3018661.3018699},
booktitle = {Proceedings of the Tenth ACM International Conference on Web Search and Data Mining},
pages = {781–789},
numpages = {9},
location = {Cambridge, United Kingdom},
series = {WSDM '17}
}

@inproceedings{RecommendationsasTreatmentsSchnabel2016,
author = {Schnabel, Tobias and Swaminathan, Adith and Singh, Ashudeep and Chandak, Navin and Joachims, Thorsten},
title = {Recommendations as treatments: debiasing learning and evaluation},
year = {2016},
publisher = {JMLR.org},
booktitle = {Proceedings of the 33rd International Conference on International Conference on Machine Learning - Volume 48},
pages = {1670–1679},
numpages = {10},
location = {New York, NY, USA},
series = {ICML'16}
}

@inproceedings{ControllingPopularityAbdollahpouri2017,
author = {Abdollahpouri, Himan and Burke, Robin and Mobasher, Bamshad},
title = {Controlling Popularity Bias in Learning-to-Rank Recommendation},
year = {2017},
isbn = {9781450346528},
publisher = {Association for Computing Machinery},
address = {New York, NY, USA},
url = {https://doi.org/10.1145/3109859.3109912},
doi = {10.1145/3109859.3109912},
booktitle = {Proceedings of the Eleventh ACM Conference on Recommender Systems},
pages = {42–46},
numpages = {5},
location = {Como, Italy},
series = {RecSys '17}
}

@inproceedings{DeconfoundedRecommendationWang2021,
author = {Wang, Wenjie and Feng, Fuli and He, Xiangnan and Wang, Xiang and Chua, Tat-Seng},
title = {Deconfounded Recommendation for Alleviating Bias Amplification},
year = {2021},
isbn = {9781450383325},
publisher = {Association for Computing Machinery},
address = {New York, NY, USA},
url = {https://doi.org/10.1145/3447548.3467249},
doi = {10.1145/3447548.3467249},
booktitle = {Proceedings of the 27th ACM SIGKDD Conference on Knowledge Discovery \& Data Mining},
pages = {1717–1725},
numpages = {9},
location = {Virtual Event, Singapore},
series = {KDD '21}
}

@inproceedings{CausalInterventionZhang2021,
author = {Zhang, Yang and Feng, Fuli and He, Xiangnan and Wei, Tianxin and Song, Chonggang and Ling, Guohui and Zhang, Yongdong},
title = {Causal Intervention for Leveraging Popularity Bias in Recommendation},
year = {2021},
isbn = {9781450380379},
publisher = {Association for Computing Machinery},
address = {New York, NY, USA},
url = {https://doi.org/10.1145/3404835.3462875},
doi = {10.1145/3404835.3462875},
booktitle = {Proceedings of the 44th International ACM SIGIR Conference on Research and Development in Information Retrieval},
pages = {11–20},
numpages = {10},
location = {Virtual Event, Canada},
series = {SIGIR '21}
}

@inproceedings{CausalEmbeddingsBonner2018,
author = {Bonner, Stephen and Vasile, Flavian},
title = {Causal embeddings for recommendation},
year = {2018},
isbn = {9781450359016},
publisher = {Association for Computing Machinery},
address = {New York, NY, USA},
url = {https://doi.org/10.1145/3240323.3240360},
doi = {10.1145/3240323.3240360},
booktitle = {Proceedings of the 12th ACM Conference on Recommender Systems},
pages = {104–112},
numpages = {9},
location = {Vancouver, British Columbia, Canada},
series = {RecSys '18}
}

@inproceedings{ModelAgnosticWei2021,
author = {Wei, Tianxin and Feng, Fuli and Chen, Jiawei and Wu, Ziwei and Yi, Jinfeng and He, Xiangnan},
title = {Model-Agnostic Counterfactual Reasoning for Eliminating Popularity Bias in Recommender System},
year = {2021},
isbn = {9781450383325},
publisher = {Association for Computing Machinery},
address = {New York, NY, USA},
url = {https://doi.org/10.1145/3447548.3467289},
doi = {10.1145/3447548.3467289},
booktitle = {Proceedings of the 27th ACM SIGKDD Conference on Knowledge Discovery \& Data Mining},
pages = {1791–1800},
numpages = {10},
location = {Virtual Event, Singapore},
series = {KDD '21}
}

@inproceedings{DisentanglingUserZheng2021,
author = {Zheng, Yu and Gao, Chen and Li, Xiang and He, Xiangnan and Li, Yong and Jin, Depeng},
title = {Disentangling User Interest and Conformity for Recommendation with Causal Embedding},
year = {2021},
isbn = {9781450383127},
publisher = {Association for Computing Machinery},
address = {New York, NY, USA},
url = {https://doi.org/10.1145/3442381.3449788},
doi = {10.1145/3442381.3449788},
booktitle = {Proceedings of the Web Conference 2021},
pages = {2980–2991},
numpages = {12},
location = {Ljubljana, Slovenia},
series = {WWW '21}
}

@inproceedings{LightGCNHe2020,
author = {He, Xiangnan and Deng, Kuan and Wang, Xiang and Li, Yan and Zhang, YongDong and Wang, Meng},
title = {LightGCN: Simplifying and Powering Graph Convolution Network for Recommendation},
year = {2020},
isbn = {9781450380164},
publisher = {Association for Computing Machinery},
address = {New York, NY, USA},
url = {https://doi.org/10.1145/3397271.3401063},
doi = {10.1145/3397271.3401063},
booktitle = {Proceedings of the 43rd International ACM SIGIR Conference on Research and Development in Information Retrieval},
pages = {639–648},
numpages = {10},
location = {Virtual Event, China},
series = {SIGIR '20}
}

@inproceedings{SelfSupervisedWu2021,
author = {Wu, Jiancan and Wang, Xiang and Feng, Fuli and He, Xiangnan and Chen, Liang and Lian, Jianxun and Xie, Xing},
title = {Self-supervised Graph Learning for Recommendation},
year = {2021},
isbn = {9781450380379},
publisher = {Association for Computing Machinery},
address = {New York, NY, USA},
url = {https://doi.org/10.1145/3404835.3462862},
doi = {10.1145/3404835.3462862},
booktitle = {Proceedings of the 44th International ACM SIGIR Conference on Research and Development in Information Retrieval},
pages = {726–735},
numpages = {10},
location = {Virtual Event, Canada},
series = {SIGIR '21}
}

@inproceedings{AreGraphAugmentationsYu2022,
author = {Yu, Junliang and Yin, Hongzhi and Xia, Xin and Chen, Tong and Cui, Lizhen and Nguyen, Quoc Viet Hung},
title = {Are Graph Augmentations Necessary? Simple Graph Contrastive Learning for Recommendation},
year = {2022},
isbn = {9781450387323},
publisher = {Association for Computing Machinery},
address = {New York, NY, USA},
url = {https://doi.org/10.1145/3477495.3531937},
doi = {10.1145/3477495.3531937},
booktitle = {Proceedings of the 45th International ACM SIGIR Conference on Research and Development in Information Retrieval},
pages = {1294–1303},
numpages = {10},
location = {Madrid, Spain},
series = {SIGIR '22}
}

@inproceedings{UnderstandingContrastiveWang2020,
author = {Wang, Tongzhou and Isola, Phillip},
title = {Understanding contrastive representation learning through alignment and uniformity on the hypersphere},
year = {2020},
publisher = {JMLR.org},
booktitle = {Proceedings of the 37th International Conference on Machine Learning},
articleno = {921},
numpages = {11},
series = {ICML'20}
}

@inproceedings{TowardsRepresentationWang2022,
author = {Wang, Chenyang and Yu, Yuanqing and Ma, Weizhi and Zhang, Min and Chen, Chong and Liu, Yiqun and Ma, Shaoping},
title = {Towards Representation Alignment and Uniformity in Collaborative Filtering},
year = {2022},
isbn = {9781450393850},
publisher = {Association for Computing Machinery},
address = {New York, NY, USA},
url = {https://doi.org/10.1145/3534678.3539253},
doi = {10.1145/3534678.3539253},
booktitle = {Proceedings of the 28th ACM SIGKDD Conference on Knowledge Discovery and Data Mining},
pages = {1816–1825},
numpages = {10},
location = {Washington DC, USA},
series = {KDD '22}
}

@inproceedings{ImprovingRecommendationChen2023,
author = {Chen, Lei and Wu, Le and Zhang, Kun and Hong, Richang and Lian, Defu and Zhang, Zhiqiang and Zhou, Jun and Wang, Meng},
title = {Improving Recommendation Fairness via Data Augmentation},
year = {2023},
isbn = {9781450394161},
publisher = {Association for Computing Machinery},
address = {New York, NY, USA},
url = {https://doi.org/10.1145/3543507.3583341},
doi = {10.1145/3543507.3583341},
booktitle = {Proceedings of the ACM Web Conference 2023},
pages = {1012–1020},
numpages = {9},
location = {Austin, TX, USA},
series = {WWW '23}
}

@ARTICLE{ASurveyonAccuracyWu2023,
author={Wu, Le and He, Xiangnan and Wang, Xiang and Zhang, Kun and Wang, Meng},
journal={ IEEE Transactions on Knowledge \& Data Engineering },
title={{ A Survey on Accuracy-Oriented Neural Recommendation: From Collaborative Filtering to Information-Rich Recommendation }},
year={2023},
volume={35},
number={05},
ISSN={1558-2191},
pages={4425-4445},
doi={10.1109/TKDE.2022.3145690},
url = {https://doi.ieeecomputersociety.org/10.1109/TKDE.2022.3145690},
publisher={IEEE Computer Society},
address={Los Alamitos, CA, USA},
month=may}

@article{GraphNeuralWu2023,
author = {Wu, Shiwen and Sun, Fei and Zhang, Wentao and Xie, Xu and Cui, Bin},
title = {Graph Neural Networks in Recommender Systems: A Survey},
year = {2022},
issue_date = {May 2023},
publisher = {Association for Computing Machinery},
address = {New York, NY, USA},
volume = {55},
number = {5},
issn = {0360-0300},
url = {https://doi.org/10.1145/3535101},
doi = {10.1145/3535101},
journal = {ACM Comput. Surv.},
month = dec,
articleno = {97},
numpages = {37}
}

@article{BiasandDebiasChen2023,
author = {Chen, Jiawei and Dong, Hande and Wang, Xiang and Feng, Fuli and Wang, Meng and He, Xiangnan},
title = {Bias and Debias in Recommender System: A Survey and Future Directions},
year = {2023},
issue_date = {July 2023},
publisher = {Association for Computing Machinery},
address = {New York, NY, USA},
volume = {41},
number = {3},
issn = {1046-8188},
url = {https://doi.org/10.1145/3564284},
doi = {10.1145/3564284},
journal = {ACM Trans. Inf. Syst.},
month = feb,
articleno = {67},
numpages = {39}
}

@inproceedings{ItemSideJiang2024,
author = {Jiang, Meng and Bao, Keqin and Zhang, Jizhi and Wang, Wenjie and Yang, Zhengyi and Feng, Fuli and He, Xiangnan},
title = {Item-side Fairness of Large Language Model-based Recommendation System},
year = {2024},
isbn = {9798400701719},
publisher = {Association for Computing Machinery},
address = {New York, NY, USA},
url = {https://doi.org/10.1145/3589334.3648158},
doi = {10.1145/3589334.3648158},
booktitle = {Proceedings of the ACM Web Conference 2024},
pages = {4717–4726},
numpages = {10},
location = {Singapore, Singapore},
series = {WWW '24}
}

@article{FairCFShao2022,
  title={FairCF: fairness-aware collaborative filtering},
  author={Shao, Pengyang and Wu, Le and Chen, Lei and Zhang, Kun and Wang, Meng},
  journal={Science China Information Sciences},
  volume={65},
  number={12},
  pages={1--15},
  year={2022},
  publisher={Springer}
}

@article{ASurveyontheFairnesWang2023,
author = {Wang, Yifan and Ma, Weizhi and Zhang, Min and Liu, Yiqun and Ma, Shaoping},
title = {A Survey on the Fairness of Recommender Systems},
year = {2023},
issue_date = {July 2023},
publisher = {Association for Computing Machinery},
address = {New York, NY, USA},
volume = {41},
number = {3},
issn = {1046-8188},
url = {https://doi.org/10.1145/3547333},
doi = {10.1145/3547333},
journal = {ACM Trans. Inf. Syst.},
month = feb,
articleno = {52},
numpages = {43}
}

@inproceedings{PopularityBiasZhu2021,
author = {Zhu, Ziwei and He, Yun and Zhao, Xing and Caverlee, James},
title = {Popularity Bias in Dynamic Recommendation},
year = {2021},
isbn = {9781450383325},
publisher = {Association for Computing Machinery},
address = {New York, NY, USA},
url = {https://doi.org/10.1145/3447548.3467376},
doi = {10.1145/3447548.3467376},
booktitle = {Proceedings of the 27th ACM SIGKDD Conference on Knowledge Discovery \& Data Mining},
pages = {2439–2449},
numpages = {11},
location = {Virtual Event, Singapore},
series = {KDD '21}
}

@inproceedings{BPRRendle2009,
author = {Rendle, Steffen and Freudenthaler, Christoph and Gantner, Zeno and Schmidt-Thieme, Lars},
title = {BPR: Bayesian personalized ranking from implicit feedback},
year = {2009},
isbn = {9780974903958},
publisher = {AUAI Press},
address = {Arlington, Virginia, USA},
booktitle = {Proceedings of the Twenty-Fifth Conference on Uncertainty in Artificial Intelligence},
pages = {452–461},
numpages = {10},
location = {Montreal, Quebec, Canada},
series = {UAI '09}
}

@inproceedings{DisentangledGraphWang2020,
author = {Wang, Xiang and Jin, Hongye and Zhang, An and He, Xiangnan and Xu, Tong and Chua, Tat-Seng},
title = {Disentangled Graph Collaborative Filtering},
year = {2020},
isbn = {9781450380164},
publisher = {Association for Computing Machinery},
address = {New York, NY, USA},
url = {https://doi.org/10.1145/3397271.3401137},
doi = {10.1145/3397271.3401137},
booktitle = {Proceedings of the 43rd International ACM SIGIR Conference on Research and Development in Information Retrieval},
pages = {1001–1010},
numpages = {10},
location = {Virtual Event, China},
series = {SIGIR '20}
}

@inproceedings{ImprovingGraphLin2022,
author = {Lin, Zihan and Tian, Changxin and Hou, Yupeng and Zhao, Wayne Xin},
title = {Improving Graph Collaborative Filtering with Neighborhood-enriched Contrastive Learning},
year = {2022},
isbn = {9781450390965},
publisher = {Association for Computing Machinery},
address = {New York, NY, USA},
url = {https://doi.org/10.1145/3485447.3512104},
doi = {10.1145/3485447.3512104},
booktitle = {Proceedings of the ACM Web Conference 2022},
pages = {2320–2329},
numpages = {10},
location = {Virtual Event, Lyon, France},
series = {WWW '22}
}

@inproceedings{UnveilingContrastiveZhang2025,
title = {Unveiling Contrastive Learning‘ Capability of Neighborhood Aggregation for Collaborative Filtering},
author={Zhang, Yu and Zhang, Yiwen and Zhang, Yi and Sang, Lei and Yang, Yun},
booktitle = {Proceedings of the 48th International ACM SIGIR Conference on Research and Development in Information Retrieval},
doi={10.1145/3726302.3730111},
pages = {1985–1994},
numpages = {10},
year={2025},
}

@inproceedings{PopularityAwareCai2024,
author = {Cai, Miaomiao and Chen, Lei and Wang, Yifan and Bai, Haoyue and Sun, Peijie and Wu, Le and Zhang, Min and Wang, Meng},
title = {Popularity-Aware Alignment and Contrast for Mitigating Popularity Bias},
year = {2024},
isbn = {9798400704901},
publisher = {Association for Computing Machinery},
address = {New York, NY, USA},
url = {https://doi.org/10.1145/3637528.3671824},
doi = {10.1145/3637528.3671824},
booktitle = {Proceedings of the 30th ACM SIGKDD Conference on Knowledge Discovery and Data Mining},
pages = {187–198},
numpages = {12},
location = {Barcelona, Spain},
series = {KDD '24}
}

@inproceedings{recbole[1.0],
author    = {Wayne Xin Zhao and Shanlei Mu and Yupeng Hou and Zihan Lin and Yushuo Chen and Xingyu Pan and Kaiyuan Li and Yujie Lu and Hui Wang and Changxin Tian and Yingqian Min and Zhichao Feng and Xinyan Fan and Xu Chen and Pengfei Wang and Wendi Ji and Yaliang Li and Xiaoling Wang and Ji{-}Rong Wen},
title     = {RecBole: Towards a Unified, Comprehensive and Efficient Framework for Recommendation Algorithms},
booktitle = {{CIKM}},
pages     = {4653--4664},
publisher = {{ACM}},
year      = {2021}
}

@inproceedings{recbole[2.0],
author    = {Wayne Xin Zhao and Yupeng Hou and Xingyu Pan and Chen Yang and Zeyu Zhang and Zihan Lin and Jingsen Zhang and Shuqing Bian and Jiakai Tang and Wenqi Sun and Yushuo Chen and Lanling Xu and Gaowei Zhang and Zhen Tian and Changxin Tian and Shanlei Mu and Xinyan Fan and Xu Chen and Ji{-}Rong Wen},
title     = {RecBole 2.0: Towards a More Up-to-Date Recommendation Library},
booktitle = {{CIKM}},
pages     = {4722--4726},
publisher = {{ACM}},
year      = {2022}
}

@inproceedings{recbole[1.2.1],
author    = {Lanling Xu and Zhen Tian and Gaowei Zhang and Junjie Zhang and Lei Wang and Bowen Zheng and Yifan Li and Jiakai Tang and Zeyu Zhang and Yupeng Hou and Xingyu Pan and Wayne Xin Zhao and Xu Chen and Ji{-}Rong Wen},
title     = {Towards a More User-Friendly and Easy-to-Use Benchmark Library for Recommender Systems},
booktitle = {{SIGIR}},
pages     = {2837--2847},
publisher = {{ACM}},
year      = {2023}
}

@inproceedings{wang2018confidence,
title={Confidence-aware matrix factorization for recommender systems},
author={Wang, Chao and Liu, Qi and Wu, Runze and Chen, Enhong and Liu, Chuanren and Huang, Xunpeng and Huang, Zhenya},
booktitle={Proceedings of the AAAI Conference on artificial intelligence},
volume={32},
number={1},
year={2018}
}

@inproceedings{wang2020setrank,
title={Setrank: A setwise bayesian approach for collaborative ranking from implicit feedback},
author={Wang, Chao and Zhu, Hengshu and Zhu, Chen and Qin, Chuan and Xiong, Hui},
booktitle={Proceedings of the aaai conference on artificial intelligence},
volume={34},
number={04},
pages={6127--6136},
year={2020}
}

@inproceedings{wang2021variable,
title={Variable interval time sequence modeling for career trajectory prediction: Deep collaborative perspective},
author={Wang, Chao and Zhu, Hengshu and Hao, Qiming and Xiao, Keli and Xiong, Hui},
booktitle={Proceedings of the Web Conference 2021},
pages={612--623},
year={2021}
}

@article{wang2021personalized,
title={Personalized and explainable employee training course recommendations: A bayesian variational approach},
author={Wang, Chao and Zhu, Hengshu and Wang, Peng and Zhu, Chen and Zhang, Xi and Chen, Enhong and Xiong, Hui},
journal={ACM Transactions on Information Systems (TOIS)},
volume={40},
number={4},
pages={1--32},
year={2021},
publisher={ACM New York, NY}
}

@inproceedings{zhu2024graph,
title={Graph signal diffusion model for collaborative filtering},
author={Zhu, Yunqin and Wang, Chao and Zhang, Qi and Xiong, Hui},
booktitle={Proceedings of the 47th International ACM SIGIR Conference on Research and Development in Information Retrieval},
pages={1380--1390},
year={2024}
}

@article{wang2025face,
  title={FACE: A General Framework for Mapping Collaborative Filtering Embeddings into LLM Tokens},
  author={Wang, Chao and Song, Yixin and Ye, Jinhui and Qin, Chuan and Shen, Dazhong and Liu, Lingfeng and Wang, Xiang and Zhang, Yanyong},
  journal={arXiv preprint arXiv:2510.15729},
  year={2025}
}

@inproceedings{10.1145/3711896.3737068,
author = {Loveland, Donald and Ju, Mingxuan and Zhao, Tong and Shah, Neil and Koutra, Danai},
title = {On the Role of Weight Decay in Collaborative Filtering: A Popularity Perspective},
year = {2025},
isbn = {9798400714542},
publisher = {Association for Computing Machinery},
address = {New York, NY, USA},
url = {https://doi.org/10.1145/3711896.3737068},
doi = {10.1145/3711896.3737068},
booktitle = {Proceedings of the 31st ACM SIGKDD Conference on Knowledge Discovery and Data Mining V.2},
pages = {1975–1986},
numpages = {12},
location = {Toronto ON, Canada},
series = {KDD '25}
}

@inproceedings{TPAB,
author = {Yoo, Hyunsik and Qiu, Ruizhong and Xu, Charlie and Wang, Fei and Tong, Hanghang},
title = {Generalizable Recommender System During Temporal Popularity Distribution Shifts},
year = {2025},
isbn = {9798400712456},
publisher = {Association for Computing Machinery},
address = {New York, NY, USA},
url = {https://doi.org/10.1145/3690624.3709299},
doi = {10.1145/3690624.3709299},
booktitle = {Proceedings of the 31st ACM SIGKDD Conference on Knowledge Discovery and Data Mining V.1},
pages = {1833–1843},
numpages = {11},
location = {Toronto ON, Canada},
series = {KDD '25}
}

@inproceedings{DCCL,
author = {Zhao, Weiqi and Tang, Dian and Chen, Xin and Lv, Dawei and Ou, Daoli and Li, Biao and Jiang, Peng and Gai, Kun},
title = {Disentangled Causal Embedding With Contrastive Learning For Recommender System},
year = {2023},
isbn = {9781450394192},
publisher = {Association for Computing Machinery},
address = {New York, NY, USA},
url = {https://doi.org/10.1145/3543873.3584637},
doi = {10.1145/3543873.3584637},
booktitle = {Companion Proceedings of the ACM Web Conference 2023},
pages = {406–410},
numpages = {5},
location = {Austin, TX, USA},
series = {WWW '23 Companion}
}


\begin{table*}[htbp]
\centering
\caption{Table of Notations}
\label{tab:notation}
\begin{tabular}{c l}
\hline
\textbf{Symbol} & \textbf{Description} \\
\hline
$\setU$ & The finite set of all users, $u \in \setU$. \\
$\setI$ & The finite set of all items, $i \in \setI$. \\
$\setS$ & The set of observed user-item interactions, $\setS \subseteq \setU \times \setI$. \\
$\setI_u^+$ & The set of items user $u$ has interacted with, $\{i \in \setI \mid (u,i) \in \setS\}$. \\
$\setU_i^+$ & The set of users who have interacted with item $i$, $\{u \in \setU \mid (u,i) \in \setS\}$. \\
$d \in \mathbb{N}$ & The dimensionality of the embedding space. \\
$\vu \in \R^d$ & The embedding vector for user $u$. \\
$\vi \in \R^d$ & The embedding vector for item $i$. \\
$\hat{r}_{ui}$ & The predicted score for a user-item pair $(u,i)$. \\
$\sigma(x)$ & The logistic sigmoid function, $1/(1+e^{-x})$. \\
\hline
\end{tabular}
\end{table*}

\appendix



\section{Detailed Theoretical Framework and Proofs}
\label{sec:appendix_theory}
This appendix provides the complete theoretical analysis, including all definitions, assumptions, and proofs that support the claims made in the main paper. A complete summary of our notation is provided in Table~\ref{tab:notation}.

\subsection{Proof of Geometric Imprint of Popularity}
\label{sec:appendix_proof_emergence}
We now provide a more formal proof for the emergence of the popularity direction, as discussed in Section~\ref{sec:imprint_of_popularity}.

\begin{assumption}[Non-zero Empirical Mean] \label{assump:init_bias}
Post-initialization, the set of user embeddings exhibits a non-zero empirical mean vector due to finite sampling variance, denoted as $\vecv = \E_{u \in \setU}[\vu^{(0)}] \neq \mathbf{0}$. We further assume the direction of this empirical mean vector remains relatively stable during the early stages of training.
\end{assumption}

\begin{proposition}[Popularity and User Distribution] \label{prop:pop_dist}
By the Law of Large Numbers, the average embedding of users who interact with an item $i$ converges to the global average user embedding as the item's popularity grows.
\begin{equation}
\lim_{Pop(i) \to \infty} \E_{u \in \setU_i^+}[\vu] = \E_{u \in \setU}[\vu] = \vecv
\end{equation}
For any sufficiently popular item $i_{pop}$, we can therefore approximate $\E_{u \in \setU_{i_{pop}}^+}[\vu] \approx \vecv$.
\end{proposition}

\begin{lemma}[Expected Gradient for Items] \label{lemma:grad_exp_item}
The gradient of the BPR loss for a triplet $(u,i,j)$ with respect to item embeddings is $\nabla_{\vi} \ell_{uij} = -\sigma(-s_{uij})\vu$ and $\nabla_{\vj} \ell_{uij} = \sigma(-s_{uij})\vu$, where $s_{uij}$ is the score margin. In early training, $\sigma(\cdot) \approx 0.5$. The total expected gradient for an item $i$ is dominated by its role as a positive sample.
\end{lemma}
\begin{proof}
The total gradient for $\vi$ is a sum of its updates as a positive item and as a negative item.
\begin{equation}
\nabla_{\vi} = \underbrace{-\sum_{u \in \setU_i^+} \E_{j}[\sigma(-s_{uij})\vu]}_{\text{Positive Role}} + \underbrace{\sum_{(u',k) \in \setS, j=i} \sigma(-s_{u'ki})\mathbf{e}_{u'}}_{\text{Negative Role}}
\end{equation}
The magnitude of the positive role term is proportional to $Pop(i)$. The magnitude of the negative role term is proportional to how often $i$ is sampled as negative, which is roughly constant for all items under uniform sampling. For a popular item, $Pop(i)$ is large, so the positive role gradient dominates. Thus, the expected gradient early in training is:
\begin{equation}
\E[\nabla_{{\vi}_{pop}}] \approx -0.5 \sum_{u \in \setU_{i_{pop}}^+} \vu = -0.5 \cdot Pop(i_{pop}) \cdot \E_{u \in \setU_{i_{pop}}^+}[\vu]
\end{equation}
Using Proposition~\ref{prop:pop_dist} and Assumption~\ref{assump:init_bias}, this simplifies to:
\begin{equation}
\E[\nabla_{{\vi}_{pop}}] \approx -0.5 \cdot Pop(i_{pop}) \cdot \vecv.
\end{equation}
This shows the expected gradient for a popular item is aligned with the mean user vector $\vecv$ and its magnitude is proportional to its popularity.
\end{proof}

\subsection{Proof of BPR Gradient Misalignment}
\label{sec:appendix_proof_misalignment}
Here we formalize the argument from Section 2.2 that the BPR gradient for a user $\vu$ is suboptimal.

\begin{definition}[Ideal Update Direction]
The Ideal Update Direction $\dopt$ for a user $u$ is the unit vector that maximizes the expected score margin between positive and negative items.
\begin{equation}
\dopt \defeq \frac{\gpos - \gneg}{\norm{\gpos - \gneg}},
\end{equation}
where $\gpos = \E_{i \in \setI_u^+}[\vi^*]$ is the barycenter of the user's positive items and $\gneg = \E_{j \notin \setI_u^+}[\vj^*]$ is the barycenter of their negative items, for converged item embeddings $\vi^*, \vj^*$.
\end{definition}

\begin{theorem}[BPR Gradient Misalignment]
The expected BPR gradient for a user embedding, $\nabla_{\vu} \mathcal{L}_u$, is not aligned with the ideal update direction $\dopt$.
\end{theorem}
\begin{proof}
The BPR gradient for user $u$ is 
$$\nabla_{\vu} \mathcal{L}_u \propto \E_{i,j}[\sigma(-\dots)(\vj^* - \vi^*)].$$
Let's analyze the direction of this gradient vector, $\E_{i,j}[\sigma(-\dots)(\vi^* - \vj^*)]$.
This is a weighted average of vectors $(\vi^* - \vj^*)$. The term $\dopt$ is proportional to the \textit{unweighted} average, $\E[\vi^*] - \E[\vj^*]$. The sigmoid term $\sigma(-\vu^T(\vi^*-\vj^*))$ introduces a bias. It gives more weight to pairs $(i,j)$ that are "hard" to rank (i.e., where the score margin is small or negative).

More importantly, the composition of the positive set $\setI_u^+$ contaminates the gradient. Let's decompose the positive contribution $\E_{i \in \setI_u^+}[w_{ui}\vi^*]$:
\begin{equation}
\begin{split}
\E_{i \in \setI_u^+}[w_{ui}\vi^*] &= \sum_{i_{niche} \in \setI_u^+} p(i_{niche})w_{ui_{niche}}\mathbf{v}_{niche}^* \\
&\quad + \sum_{i_{pop} \in \setI_u^+} p(i_{pop})w_{ui_{pop}}\mathbf{v}_{pop}^*.
\end{split}
\end{equation}
Even if the number of popular items is small, their embeddings $\mathbf{v}_{pop}^*$ have large magnitudes and are aligned with $\popdir$. This can cause the second term to dominate, pulling the entire gradient direction towards $\popdir$ and away from the direction defined by the user's niche preferences. The resulting gradient is a compromise between the true preference direction and the popularity direction, hence it is not aligned with the optimal direction $\dopt$. This leads to inefficient optimization as described in the main text.
\end{proof}

\subsection{Derivation of the Decoupled Optimization Framework}
\label{sec:appendix_derivation_ddc}
The analysis reveals that the BPR gradient conflates a preference signal with a popularity signal. We can achieve a better optimum by decoupling them, as outlined in Section 2.4.

\begin{theorem}[Derivation of the DDC Framework]
Let a model be trained to convergence, yielding sub-optimal user embeddings $\vuorig$. A superior solution can be found by reframing the optimization problem to learn user-specific scalar corrections $(\alpha_u, \beta_u)$ in a modified BPR loss that explicitly decouples the updates for preference alignment and popularity calibration.
\end{theorem}
\begin{proof}
The goal of an update step for user $u$ is to move the embedding $\vu$ along $\dopt \propto (\gpos - \gneg)$. This ideal update can be decomposed into two conceptual components: a pull towards the positive barycenter $\gpos$ and a push away from the negative barycenter $\gneg$. The standard BPR gradient is a poor proxy for this.

Our strategy is to freeze the base embeddings $(\vuorig, \{\vi^*\})$ and introduce learnable corrections. We operationalize the decoupling insight by assigning different functional forms to the user embedding when paired with a positive versus a negative item.

\begin{enumerate}
    \item \textbf{Positive Interaction Term}: The objective is to increase $\vu^T\vi^*$. The ideal direction for this is towards $\gpos$. We approximate this with a pre-computed, personalized preference direction ${\predir}_u \approx \gpos / \norm{\gpos}$. We modify the user embedding with a learnable component along this direction:
    \begin{equation}
    \vu^{pos-term} = \vuorig + \beta_u {\predir}_u.
    \end{equation}
    The scalar $\beta_u$ learns the optimal step size for user $u$ along their specific preference axis.
    \item \textbf{Negative Interaction Term}: The objective is to decrease $\vu^T\vj^*$. The ideal direction for this is away from $\gneg$. Since the barycenter of negative items aligns with the popularity direction (as their specific features largely cancel out), $\gneg$ is strongly aligned with $\popdir$. The ideal calibration is thus a push in the direction of $-\popdir$. We modify the user embedding with a learnable component along $\popdir$:
    \begin{equation}
    \vu^{neg-term} = \vuorig + \alpha_u \popdir.
    \end{equation}
    The scalar $\alpha_u$ learns the optimal amount of popularity calibration for user $u$. We expect $\alpha_u$ to become negative.
\end{enumerate}

By substituting these purpose-specific user embeddings into the BPR loss, we derive the DDC loss function for the user-specific parameters $(\alpha_u, \beta_u)$:
\begin{equation}
\label{eq:ddc_loss}
\mbox{\footnotesize $\displaystyle
\begin{aligned}
\mathcal{L}_{DDC} &= \sum_{(u,i,j) \in \mathcal{D}} -\ln \sigma\left( (\vu^{pos-term})^T \vi^* - (\vu^{neg-term})^T \vj^* \right) \\
&= \sum_{(u,i,j) \in \mathcal{D}} -\ln \sigma\left( (\vuorig + \beta_u {\predir}_u)^T \vi^* - (\vuorig + \alpha_u \popdir)^T \vj^* \right)
\end{aligned}
$}
\end{equation}
This is precisely the loss function from Equation~\ref{eq:ddc_loss}. The optimization finds the optimal scalar corrections $(\alpha_u^*, \beta_u^*)$. The final, improved user embedding logically incorporates both corrections:
\begin{equation}
\vu^{final} \defeq \vuorig + \alpha_u^* \popdir + \beta_u^* {\predir}_u.
\end{equation}
This framework succeeds because it resolves the gradient conflict of standard BPR. It replaces a single, confounded, high-dimensional gradient update with two targeted, one-dimensional updates along interpretable and causally relevant axes, allowing the user embedding to move to a superior position in the latent space.
\end{proof}

\balance
\end{document}